\documentclass{emulateapj}
\pdfoutput=1
\usepackage{apjfonts}
\usepackage{graphicx}
\usepackage{amssymb}
\usepackage{amsmath}
\usepackage{bm}

\received{August 13, 2020}
\accepted{November 18, 2020}

\shorttitle{YMC formation by fast H{\sc i} gas collision}
\shortauthors{R. Maeda et al.}

\begin{document}

\title{Formation of Massive Star Clusters by Fast H{\sc i} Gas Collision}
\email{maeda.ryunosuke@nagoya-u.jp}
\author{Ryunosuke Maeda}
\affiliation{Department of Physics, Graduate School of Science, Nagoya University,
Furo-cho, Chikusa-ku, Nagoya 464-8602, Japan}

\author{Tsuyoshi Inoue}
\affiliation{Department of Physics, Graduate School of Science, Nagoya University,
Furo-cho, Chikusa-ku, Nagoya 464-8602, Japan}

\author{Yasuo Fukui}
\affiliation{Department of Physics, Graduate School of Science, Nagoya University,
Furo-cho, Chikusa-ku, Nagoya 464-8602, Japan}

\begin{abstract}
Young massive clusters (YMCs) are dense aggregates of young stars, which are essential to galaxy evolution, owing to their ultraviolet radiation, stellar winds, and supernovae. The typical mass and radius of YMCs are $M\sim 10^4\ \mathrm{M_{\odot}}$ and $R \sim 1\ \mathrm{pc}$, respectively, indicating that many stars are located in a small region. The formation of YMC precursor clouds may be difficult because a very compact massive cloud should be formed before stellar feedback blows off the cloud. Recent observational studies suggest that YMCs can be formed as a consequence of the fast H{\sc i} gas collision with a velocity of $\sim 100\ \mathrm{km\ s^{-1}}$, which is the typical velocity of the galaxy-galaxy interaction. In this study, we examine whether the fast H{\sc i} gas collision triggers YMC formation using three-dimensional magnetohydrodynamics simulations, which include the effects of self-gravity, radiative cooling/heating, and chemistry. We demonstrate that massive gravitationally bound gas clumps with $M >10^4\ \mathrm{M_{\odot}}$ and $L \sim 4\ \mathrm{pc}$ are formed in the shock compressed region induced by the fast H{\sc i} gas collision, which massive gas clumps can evolve into YMCs. Our results show that the YMC precursors are formed by the global gravitational collapse of molecular clouds, and YMCs can be formed even in low-metal environments, such as the Magellanic Clouds. Additionally, the very massive YMC precursor cloud, with $M >10^5\ \mathrm{M_{\odot}}$, can be created when we consider the fast collision of H{\sc i} clouds, which may explain the origin of the very massive stellar cluster R136 system in the Large Magellanic Cloud. 

\end{abstract}


\section{INTRODUCTION} \label{sec:intro}
Young massive clusters (YMCs) are dense young star clusters with a typical mass of $\sim 10^4\ \mathrm{M_{\odot}}$ and radius of $\sim 1\ \mathrm{pc}$ \citep{portegies2010young}.
The typical density of YMCs is $> 10^3\ \mathrm{M_{\odot}\ pc^{-3}} $ \citep{longmore2014formation}, which is significantly larger than that of field stars in the solar neighborhood $0.01\ \mathrm{M_{\odot}\ pc^{-3}}$ \citep{holmberg2000local}.
Approximately one hundred YMCs have been discovered in our local group galaxies, and galaxies that contain a few thousand YMCs have also been found outside our local group galaxies \citep{longmore2014formation}.
Typical YMCs contain tens of massive stars,
suggesting that they are crucial for galaxy evolution, owing to their ultraviolet (UV) radiation, stellar winds, and supernovae.
Despite their importance, the formation mechanism of YMCs is not well understood.

\cite{Fujii2015,fujii2016formation} investigated the formation and evolution of YMCs from turbulent molecular clouds using hydrodynamics and N-body simulations.
Their results showed that a very large velocity dispersion of $\sim 20\ \mathrm{km\ s^{-1}}$ is necessary for the parent molecular cloud to evolve into the YMC.
This velocity dispersion is greater than that of typical the giant molecular clouds as expected from Larson's law \citep{larson1981turbulence}. 
Thus, their result suggested that YMCs can be formed in environments where strong shock waves and/or strong turbulence is expected, such as the sites of cloud collision, the galactic merger, and the galactic center. 
However, it remains unknown whether such environments can induce YMC precursor clouds.

Recent observational studies of gas structures around massive star clusters in the Large Magellanic Cloud (LMC) reveal new information on YMC formation.
\cite{fukui2017formation} and \cite{tsuge2019formation} conducted a detailed analysis of H{\sc i} gas surrounding YMCs in the LMC using the Australia Telescope Compact Array and Parkes H{\sc i} gas data \citep{staveley1997hi,kim1998hi,kim2003neutral}.
They discovered that there are characteristic features caused by the fast H{\sc i} gas collision, such as bridge features in the velocity structures and complementary spatial gas distributions around YMC forming regions.
The observed H{\sc i} gas collisions have a spatial scale of $\sim 1\ \mathrm{kpc}$ and the relative velocity of the collision is as high as $100\ \mathrm{km\ s^{-1}}$ \footnote{This collision velocity is derived from the line-of-sight velocity ($\sim 60 \ \mathrm{km\ s^{-1}}$) by assuming tilting angle of $\sim 45$ degrees.}.
The simulations by \cite{fujimoto1990asymmetric} and \cite{Bekki2007} showed that a gas flow from the Small Magellanic Cloud (SMC) to the LMC, which resembles the colliding flows found in the above observations, can be created, owing to a past close encounter of the LMC and SMC.
Furthermore, the metallicity measured in the gas colliding region is smaller than the metallicity at the optical stellar bar region in the LMC \citep{fukui2017formation,tsuge2019formation}, which is consistent with the H{\sc i} gas inflow scenario from the SMC. 
The results of these previous studies suggest that the fast H{\sc i} gas collision triggers YMC formation, which includes the R136 system (the most massive YMC in our local group galaxies).
Such gas collisions around the YMCs are observed in the Milky Way \citep{furukawa2009molecular,ohama2010temperature,fukui2013molecular,fukui2016two,kuwahara2019cluster,fujita2020massive} and in the outer galaxies \citep{tachihara2018triggering,tsuge2019formation2,tsuge2020formation}, which include the collision of molecular clouds. 
These studies indicate that gas collision is an important mechanism for YMC formation. 

H{\sc i} gas collision simulations have been performed in many studies, in the context of H{\sc i}/molecular cloud formation \citep{inoue2008two,inoue2009two,Inoue2016,Koyama2000,Koyama2001,ballesteros1999turbulent,hartmann2001rapid,hennebelle2008warm,banerjee2009clump,heitsch2005formation,heitsch2008cooling,heitsch2006birth,heitsch2009effects,vazquez2007molecular,vazquez2006molecular}.
These simulations employed a collision speed of $\sim 10\ \mathrm{km\ s^{-1}}$, which is typical of interstellar cloud formation sites \cite[e.g.,][]{inutsuka2015formation}. 
Thus, theoretically, it is still unclear whether the fast H{\sc i} gas collision, with an observed speed of $\sim 100\ \mathrm{km\ s^{-1}}$, can actually induce YMC formation.

In this study, we perform fast H{\sc i} gas collision simulations using a three-dimensional (3D) magnetohydrodynamics (MHD) code that includes the effects of interstellar cooling/heating, self-gravity, and chemistry.
In Section 2, we introduce the theory and simulation setup. 
In Section 3, we present the results of the simulations and analyze the results. 
In Section 4, we discuss the implications of the results and, finally, we summarize the study in Section 5.

\section{METHODS} \label{sec:Method}
\subsection{Theory}
To examine interstellar gas dynamics, we employ 3D ideal MHD simulations, including the effects of chemical reactions, radiative cooling/heating, and self-gravity. 
We solve the following equations:
\begin{equation}
\frac{\partial n_{\alpha}}{\partial t}+\nabla \cdot(n_{\alpha} \vec{v})=f_{\mathrm{\alpha}}\left(n_{\mathrm{\beta}}, N_{\mathrm{\beta}}, T, G_0 \right),
\end{equation}

\begin{align}
  \begin{split}
    \frac{\partial}{\partial t} \left(\rho \vec{v} \right)+\nabla \cdot  &\left( \rho \vec{v} \otimes \vec{v} + p -\frac{1}{4 \pi} \vec{B} \otimes \vec{B} \right. \\
    & \left. +\frac{1}{8 \pi} B^2 \right)=-\rho \nabla \Phi,
  \end{split}
\end{align}
\begin{align}
  \begin{split}
    \frac{\partial}{\partial t} (\frac{1}{2} \rho v^{2} &+\rho \epsilon+\frac{B^{2}}{8 \pi})+\nabla \cdot \left[\rho \vec{v}\left(\frac{1}{2} v^{2}+h\right) \right. \\
    & \left.+\frac{1}{4 \pi}(\vec{B} \times \vec{v}) \times \vec{B}\right]=-\rho \mathcal{L}\left(n_{\mathrm{\beta}}, N_{\mathrm{\beta}}, T, G_0\right),
  \end{split}
\end{align}
\begin{equation}
\epsilon=\frac{1}{\gamma-1} \frac{p}{\rho}\ , \ h=\frac{\gamma}{\gamma-1} \frac{p}{\rho},
\end{equation}
\begin{equation}
\frac{\partial \vec{B}}{\partial t}+\mathbf{\nabla} \times(\vec{B} \times \vec{v})=0,
\end{equation}
\begin{equation}
\nabla \cdot \vec{B}=0,
\end{equation}
\begin{equation}
\nabla^{2} \Phi=4 \pi G \rho,
\end{equation}
\begin{equation}
\rho=\sum_{\mathrm{\alpha}} m_{\mathrm{\alpha}} n_{\mathrm{\alpha}},
\end{equation}
where $\rho,\ p,\ T,\ \vec{v},$ and $\vec{B}$ are the density, pressure, temperature, velocity, and magnetic field; $\epsilon$ and $h$ are the specific internal energy and enthalpy; $\gamma$ is the specific heat ratio; the subscripts $\alpha$ and $\beta$ denote the chemical species $\mathrm{p},\ \mathrm{H},\ \mathrm{H}_{2},\ \mathrm{He},\ \mathrm{He}^{+},\ \mathrm{C},\ \mathrm{C}^{+},$ and $\mathrm{CO}$; $n_{\alpha}$ is the number density of the species $\alpha$; $N_\beta$ is the column density of the species $\beta$; $f_{\mathrm{\alpha}}$ is the chemical reaction of the species $\alpha$; $\Phi$ is the gravitational potential; $G$ is the gravitational constant; $G_0$ is the background UV field strength; and $\mathcal{L}$ is the net cooling rate per unit volume. 
We consider the realistic chemical reactions and cooling/heating processes presented in \cite{inoue2012formation}. Here, we briefly summarize the involved microphysics: Chemical reactions considered in our simulations are ionizations of $\mathrm{H}$, $\mathrm{He}$, and $\mathrm{C}$ owing to cosmic-rays \citep{millar1997umist}, $\mathrm{H}_{2}$ formation on dust grains \citep{tielens1985photodissociation}, $\mathrm{H}_{2}$ photodissociation owing to UV radiation \citep{draine1996structure}, $\mathrm{H}_{2}$ and $\mathrm{CO}$ dissociation owing to collisions with $\mathrm{e}$, $\mathrm{p}$, and $\mathrm{H}$ particles \citep{hollenbach1989molecule}, $\mathrm{H}_{2}$ and $\mathrm{CO}$ dissociation owing to collision with $\mathrm{He}^+$  \citep{millar1997umist}, ionization of $\mathrm{H}$, $\mathrm{He}$, and $\mathrm{C}$ owing to collisions with $\mathrm{e}$, $\mathrm{p}$, $\mathrm{H}$, and $\mathrm{H_2}$ particles \citep{millar1997umist,hollenbach1989molecule}, recombination of $\mathrm{H}^{+}$, $\mathrm{He}^{+}$, and $\mathrm{C}^{+}$ \citep{hollenbach1989molecule,shapiro1987hydrogen}, formation of $\mathrm{CO}$ molecules \citep{nelson1997dynamics}, photodissociation of $\mathrm{CO}$ molecules \citep{nelson1997dynamics,lee1996photodissociation}, and photoionization of $\mathrm{C}$ particles \citep{tielens1985photodissociation}; The heating/cooling processes considered in our simulations are photoelectric heating by dust drains and polycyclic aromatic hydrocarbons (PAHs) \citep{bakes1994photoelectric,wolfire2003neutral}, cosmic-ray heating \citep{goldsmith1978molecular}, $\mathrm{H}_{2}$ photodissociative heating \citep{black1977models}, cooling owing to Ly$\alpha$ line emission \citep{spitzer1978physical}, cooling owing to $\mathrm{C}^+$ line emission \citep{de1980hydrostatic}, cooling owing to $\mathrm{O}$ line emission \citep{wolfire2003neutral,de1980hydrostatic}, $\mathrm{CO}$ ro-vibrational cooling \citep{Hosokawa2006,hollenbach1989molecule,hollenbach1979molecule}, and cooling owing to the recombination of electrons with grains and PAHs \citep{bakes1994photoelectric}.
Note that the effects of UV shielding in chemical reactions and heating/cooling process follow those in \cite{inoue2012formation}. 

We solve the above set of complicated equations using the operator-splitting technique. 
The MHD part of the basic equations is solved in conservative fashion using a second-order Godunov-type finite-volume scheme \citep{van1997flux} developed by \cite{sano1999higher}. 
The induction equation is solved using the consistent method of characteristics with constrained transport \citep{clarke1996consistent}. 
The cooling/heating terms are solved by employing a second-order explicit scheme; for chemical reactions, we use the piecewise exact solution method developed by \cite{inoue2008two}. 
The Poisson equation is solved using the multigrid method \citep{press1986numerical}. 

\subsection{Numerical Settings}
\begin{figure}[ht!]
\plotone{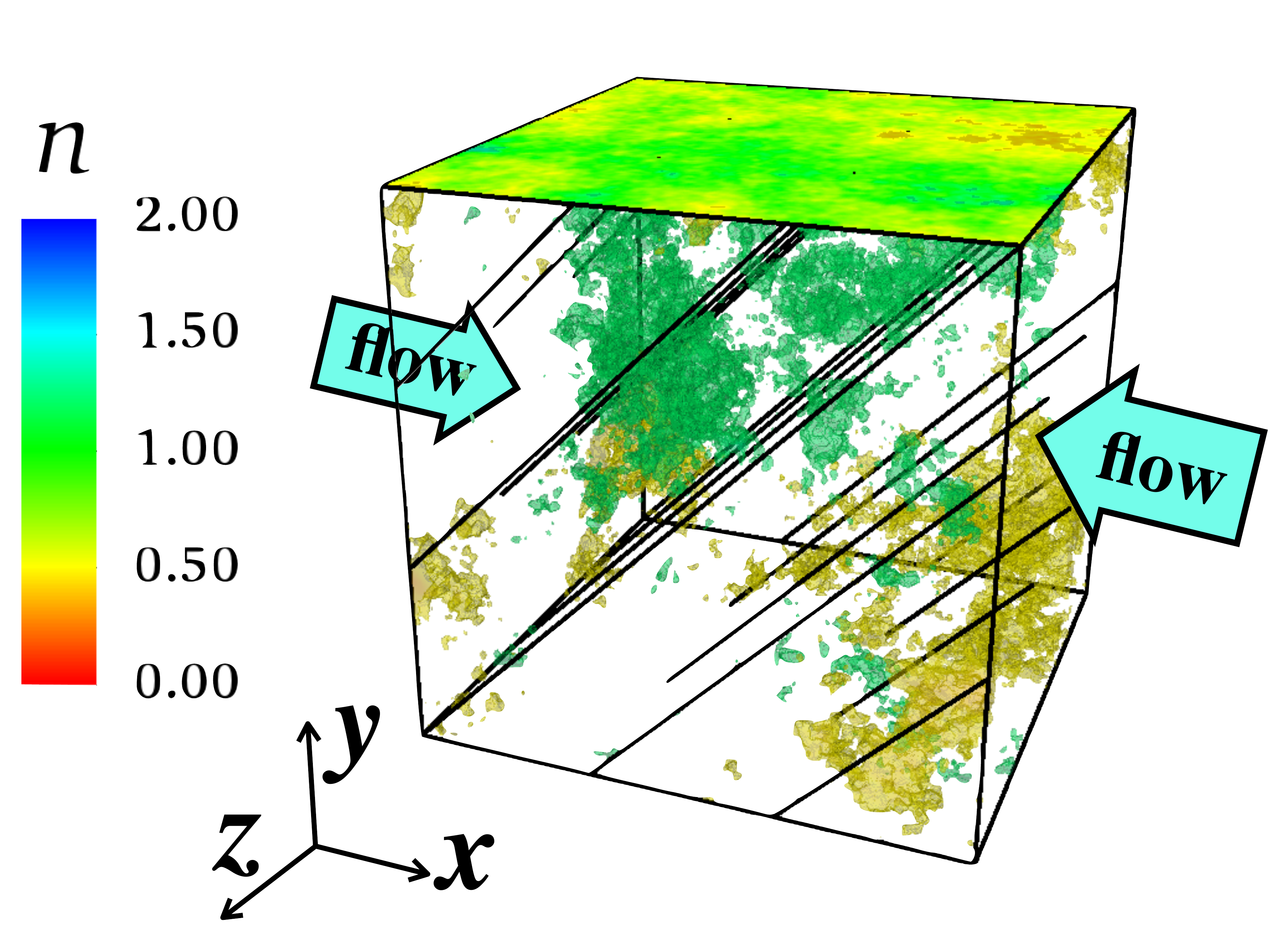}
\caption{ Schematic of the initial condition. The color indicates the initial density for the model D1B1Z1. The top panel ($y=100 \ \mathrm{pc}$) shows two-dimensional density cross section. In the box, low-density (high-density) regions due to the initial density fluctuations are shown as yellow (green) regions. The amplitude of the initial density inhomogeneity is 20\% of the initial mean density.  The black lines represent the initial magnetic field lines and the arrows indicate the orientations of the converging flows. \label{fig:initial}}
\end{figure}

As mentioned in Section 1, the scale of an H{\sc i} gas colliding region is as large as $1\ \mathrm{kpc}$, which is significantly larger than that of the YMC ($<\sim 10\ \mathrm{pc}$).
Because we are interested in the details of the YMC formation mechanism, we locally simulate the evolution of a shock compressed layer, created by the H{\sc i} gas collision, using the $L_{\mathrm{box}} = 100\ \mathrm{pc}$ scale (cubic) numerical domain. 
We divide the numerical domain into $512^3$ uniform cells, indicating that the numerical resolution is $\sim 0.2\ \mathrm{pc}$. 
The converging H{\sc i} gas flows along the $x$-axis are used to mimic the observed fast H{\sc i} gas collision in the LMC. 
The mean initial density of the gas flow is $\left\langle n_{0}\right\rangle \sim 1$ or $10\ \mathrm{cm^{-3}}$ (depending on the model), which corresponds to the density of the typical warm neutral medium (WNM) or cold neutral medium (CNM), respectively. 
We use a relative velocity of $100\ \mathrm{km\ s^{-1}}$ for the converging H{\sc i} gas flows, which is consistent with the observed value \citep{fukui2017formation,tsuge2019formation}. 
We add the fluctuations to the initial density with the Kolmogorow power spectrum (see Figure \ref{fig:initial}). 
The density dispersion is set at $20\%$ of the $\left\langle n_{0}\right\rangle$. 
The initial thermal pressure is set at $4.5\times10^3\ (1.6\times10^3)\ \mathrm{K\ cm^{-3}}$ for the $\left\langle n_{0}\right\rangle\sim 1\ (10)\ \mathrm{cm^{-3}}$ case, which yields $T= 4.5\times 10^3\ (1.6\times 10^2) \ \mathrm{K}$.
A uniform magnetic field, oriented at $45^{\circ}$ with respect to the $x$-axis in the $x$-$y$ plane, is initially set with a strength of $1.0$ or $3.0\ \mathrm{\mu G}$, depending on the model. 
This strength is consistent with observed values of the coherent magnetic field in the whole LMC \citep{gaensler2005magnetic}. 
We study two cases of the initial gas metallicity: the solar metallicity case ($x_{\mathrm{H}} \equiv n_{\mathrm{H}} / \sum n_{\mathrm{i}}=0.91$, $x_{\mathrm{p}} = 9.4 \times 10^{-3}$, $x_{\mathrm{H}_{2}}=9.4 \times 10^{-9}$, $x_{\mathrm{He}}=9.0 \times 10^{-2}$, $x_{\mathrm{He}^{+}}=5.6 \times 10^{-4}$, $x_{\mathrm{C}^{+}}=1.4 \times 10^{-4}$, $x_{\mathrm{C}}=1.6 \times 10^{-9}$, $x_{\mathrm{CO}}=2.2\times 10^{-21}$, $x_{\mathrm{O}}=3.2 \times 10^{-4}$) and one-fifth of the solar metallicity case ($x_{\mathrm{H}} =0.91$, $x_{\mathrm{p}} = 9.4 \times 10^{-3}$, $x_{\mathrm{H}_{2}}=9.4 \times 10^{-9}$, $x_{\mathrm{He}}=9.0 \times 10^{-2}$, $x_{\mathrm{He}^{+}}=5.6 \times 10^{-4}$, $x_{\mathrm{C}^{+}}=2.8 \times 10^{-5}$, $x_{\mathrm{C}}=3.2 \times 10^{-10}$, $x_{\mathrm{CO}}=4.4\times 10^{-22}$, $x_{\mathrm{O}}=6.4 \times 10^{-5}$), where $x$ indicates abundance. 
One-fifth of the solar metallicity approximates to the mean metallicity value in the LMC and SMC. 
The background far UV (FUV) intensity is set to the Habing flux ${G}_{0}=1.6 \times 10^{-3} \ \mathrm{erg}\ \mathrm{cm}^{-2}\ \mathrm{s}^{-1}$ \citep{habing1968interstellar}.

Periodic boundary conditions are imposed on the $y, z$ boundaries. For the $x$-boundaries, we impose continuous gas flows, i.e., the boundary value of a physical quantity $f$ at $x=0$ and $x=L_{\mathrm{box}}$ are $f(t, x=0, y, z)=f\left(t=0, x=L_{\mathrm{box}}-v_{\mathrm{col}} t, y, z\right)$ and $f\left(t, x=L_{\mathrm{box}}, y, z\right)=f\left(t=0, x=v_{\mathrm{col}} t, y, z\right)$, respectively, where $v_{\mathrm{col}}$ is $50\ \mathrm{km\ s^{-1}}$ in our simulations. 
The value of the gravitational potential of the $x$ boundaries is calculated using the method developed in \cite{miyama1987fragmentation}. 
We perform simulations with the set of initial parameters summarized in Table \ref{tab:models}.

\begin{deluxetable*}{lccccccc}
\tablenum{1}
\tablecaption{Parameters and results of our simulations\label{tab:models}}
\tablewidth{0pt}
\tablehead{
\colhead{Name} & \colhead{ $\left\langle n_{0}\right\rangle$ ($\mathrm{cm^{-3}}$)} & \colhead{$B_{0}$ ($\mathrm{\mu G}$)} & \colhead{$Z_0$ ($\mathrm{Z_{\odot}}$)} 
& \colhead{$M_{\mathrm{clump,max}}(\mathrm{M_{\odot}})$}  & \colhead{$L_{\mathrm{clump,max}}(\mathrm{pc})$}& \colhead{$t_{\mathrm{form}}(\mathrm{Myr})$}& \colhead{$t_{\mathrm{form}}-t_{\mathrm{mol}}(\mathrm{Myr})$}}
\startdata
D1B1Z1         & $ 0.79$    &   $1.0$     &   $1.0 $  & $\sim4 \times 10^4$ &$\sim 4$ & 24.5& 7.0\\
D1B3Z1         & $ 0.79$    &   $3.0$     &   $1.0 $ &$-$ & $-$&$-$ & $-$\\
D1B1Z0.2         & $ 0.79$    &   $1.0$     &   $0.20 $ & $\sim4 \times 10^4$ &$\sim 5$  &24.0 & 1.0 \\
D10B3Z1         & $ 10$   &   $3.0$    &   $1.0 $ &$\sim2 \times 10^5$ &$\sim 6$  & 8.0& 6.0
\enddata
\end{deluxetable*}

\section{RESULTS} \label{sec:result}
\subsection{The Results of the Fiducial Model D1B1Z1}
\begin{figure*}[!t]
\centering
\includegraphics[clip, width=\textwidth]{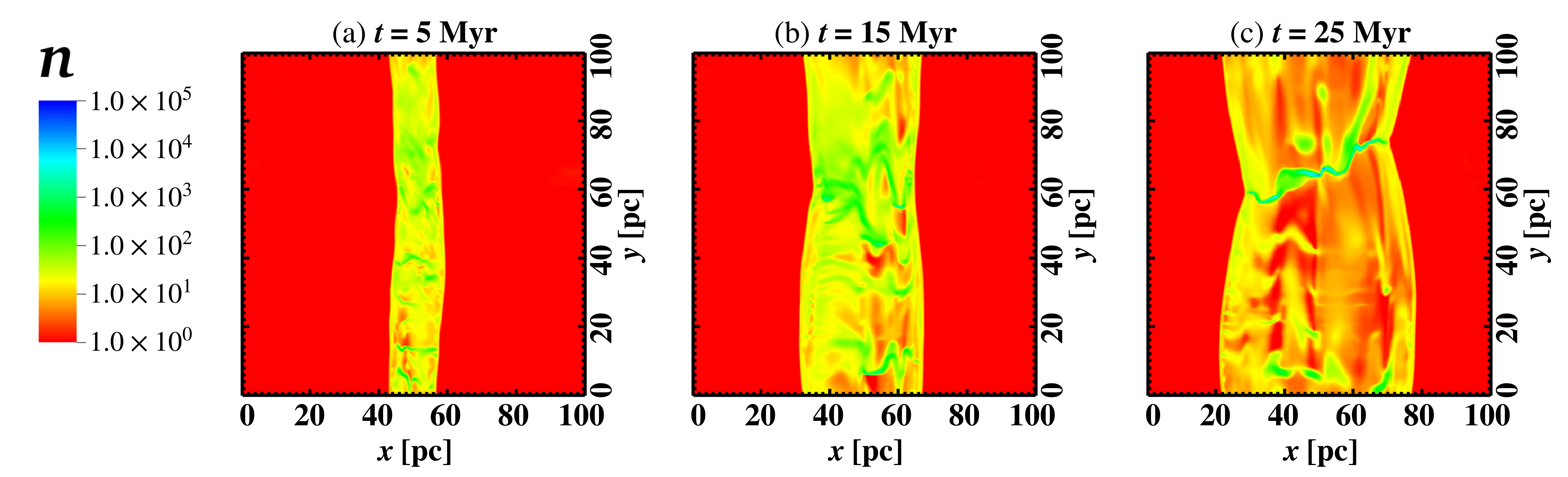}
\caption{ Two-dimensional density cross sections of the D1B1Z1 result at (a) $t=5\ \mathrm{Myr}$, (b) $15\ \mathrm{Myr}$, and (c) $25\ \mathrm{Myr}$. }
\label{fig:result-rho}
\end{figure*}

In Figure \ref{fig:result-rho}, we present the snapshots of two-dimensional density fields in the $z=0$ plane at $t=5,\ 15$, and $25\ \mathrm{Myr}$. 
We can observe the two shock waves induced by the collision. 
The typical cooling timescale of the shock compressed H{\sc i} gas layer is given by 
\begin{equation}
\begin{aligned}
t_{\text {cool}} & \simeq 0.4\ \mathrm{Myr}\\ & \left(\frac{0.2\ Z_{\odot}}{ \mathrm{Z_{\odot}}}\right)^{-1} \left(\frac{\bar{n}}{4\ \mathrm{cm^{-3}}}\right)^{-3 / 2} \left(\frac{\bar{p}/k_{\mathrm{B}}}{2\times 10^5\ \mathrm{K\ cm^{-3}}} \right)^{1 / 2},
\end{aligned} \label{eq:coolingtime}
\end{equation}
where $\bar{n}$ is the number density at the postshock region, $\bar{p}$ is the thermal pressure at the postshock region, and $k_{\mathrm{B}}$ is the Boltzmann constant \citep{Inoue2015}. 
We have estimated $\bar{n}$ and $\bar{p}$ using the adiabatic shock jump condition. 
After the cooling time, the postshock layer loses thermal pressure and begins to contract. 
This contraction amplifies the components of the magnetic field perpendicular to the converging flows.
The amplification ends when the magnetic pressure balances with the ram pressure of the converging flows \citep{inoue2008two,inoue2009two,heitsch2008cooling}. 

\begin{figure}[ht!]
\plotone{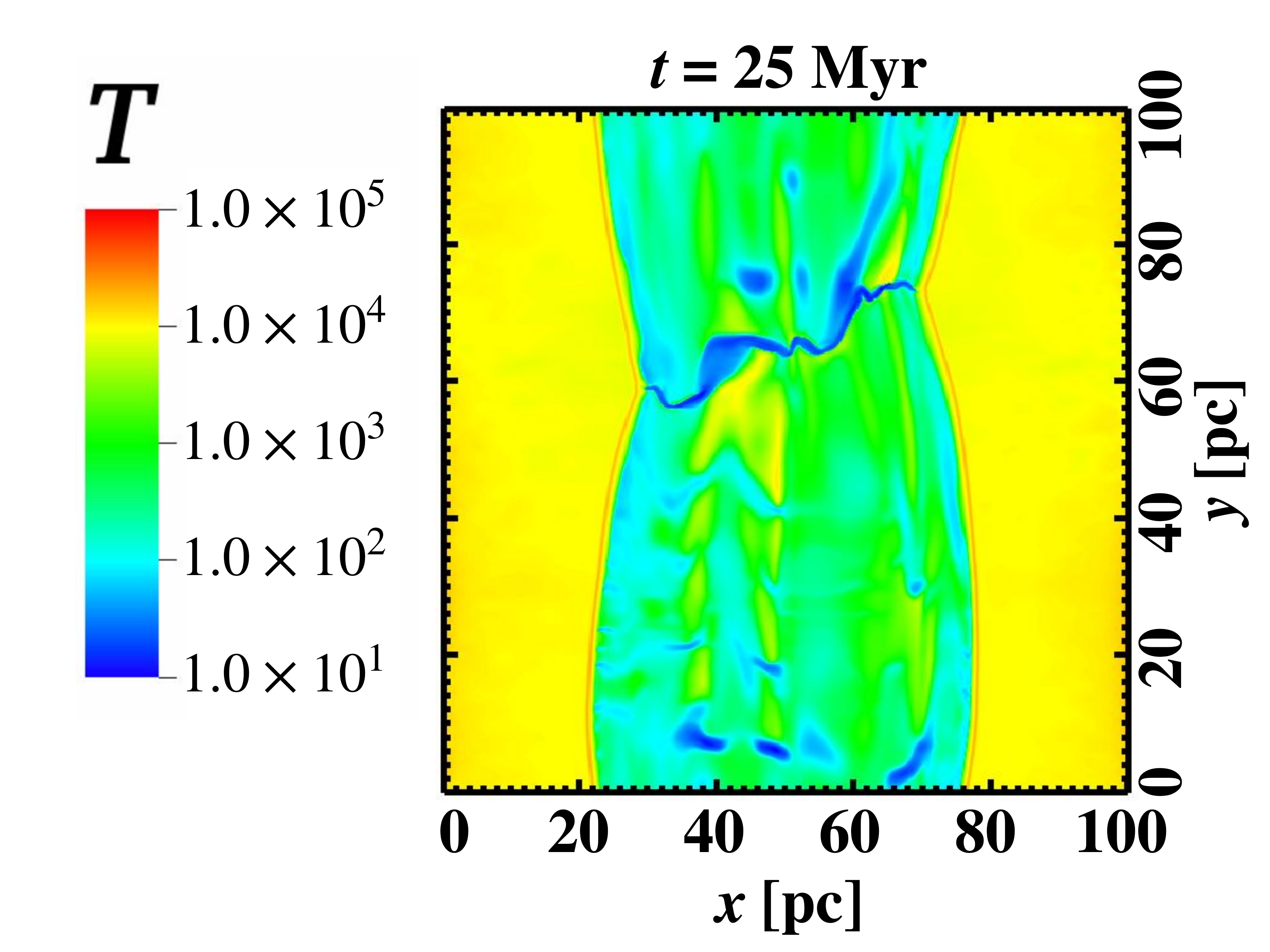}
\caption{ A two-dimensional cross section of the temperature of the D1B1Z1 result at $t=25\ \mathrm{Myr}$. \label{fig:T}}
\end{figure}

Figure \ref{fig:T} shows the temperature structure of the result of the D1B1Z1 model at $t=25\ \mathrm{Myr}$. 
We can confirm that the temperature rises at the shock fronts, and then decreases rapidly by radiative cooling. 

Because the cooling H{\sc i} gas layer is thermally unstable \citep{field1965thermal}, depending on the metallicity and FUV strength \citep{Inoue2015}, 
the postshock gas evolves into a two-phase medium, composed of WNM and CNM. 
The formation of the two-phase medium is confirmed in Figure \ref{fig:result-rho} and  \ref{fig:T}. 
 
Because the thermal instability grows along the magnetic field line and the created CNM regions are stretched by postshock turbulent flows, the CNM has a highly filamentary morphology \citep{Inoue2016}. 

The free-fall time of a gas sheet can be estimated as 
\begin{equation}
    t_{\mathrm{ff, sheet}}=\frac{1}{\sqrt{2 \pi G \bar{\rho}}}\sim 6\ \mathrm{Myr}\left(\frac{\bar{\rho}}{35\ \mathrm{m_p}\  \mathrm{cm}^{-3}}\right)^{-\frac{1}{2}}  , \label{eq:tff}
\end{equation}
where $\bar{\rho}$ is the average shocked gas density and $\mathrm{m_p}$ is the mass of a proton. 
From \cite{inoue2009two}, $\bar{\rho}$ is given by 
\begin{equation}
\bar{\rho} \sim 35\ \mathrm{m_p}\ \mathrm{cm}^{-3}\left(\frac{v_{\mathrm{sh}}}{50\ \mathrm{km}\ \mathrm{s}^{-1}}\right)\left(\frac{\rho_{0}}{1\ \mathrm{m_p}\  \mathrm{cm}^{-3}}\right)^{\frac{3}{2}}\left(\frac{B_{0,\perp}}{1\ \mu \mathrm{G}}\right)^{-1},  \label{eq:shockjump-n}
\end{equation}
where $\rho_{\mathrm{0}}$ is the density of the preshock region, $B_{\mathrm{0},\perp}$ is the magnetic field strength of the preshock region perpendicular to the converging flows, and $v_{\mathrm{sh}}$ is the shock velocity. 
This free-fall time provides an approximate timescale after which self-gravity starts to influence the postshock region. 
In the present case, the postshock CNM regions are gathered by gravity, and we can finally observe the formation of very compact and dense gas clumps (blue regions) in Panel (c) of Figure \ref{fig:result-rho}. 
In the following subsection, we will identify the dense gas clumps and examine whether they can evolve into a YMC.

\subsection{Identification of YMC Precursor Candidates}
To form the YMC, the precursor gas cloud should be massive and compact $(M\gtrsim 10^4\ \mathrm{M_{\odot}}$ and $L\lesssim 10\ \mathrm{pc})$ before it consumes the cloud. 
We identify the candidates of the YMC forming regions by employing the following procedure. 

First, we estimate the best snapshot time at which the compact star-forming gas cloud is already formed, but not many gas is consumed by star formation. 
Because we do not explicitly solve the star formation, we estimate the mass of stars by computing the following amounts, assuming star formation efficiency (SFE): 
\begin{equation}
M_{\mathrm{star}}(t)=\int^t \mathrm{SFR}(t')\ dt', \label{eq:mstar}
\end{equation}
\begin{equation}
    \mathrm{SFR}(t)=\int \epsilon_{\mathrm{SFE}} \frac{\rho(t)}{t_{\mathrm{ff}}(t)} \ dV,
\end{equation}
where $\epsilon_{\mathrm{SFE}}$ is SFE and $t_{\mathrm{ff}}$ is the local free-fall time. 
Here, we adopt $\epsilon_{\mathrm{SFE}}=0.02$, which is consistent with the observed value \citep[e.g.,][]{krumholz2011universal}. 
In this study, we use snapshot data, at which $10\%$ of the postshock gas mass is converted into the stars, i.e., at the time 
of $M_{\mathrm{star}} (t) = 0.1 M_{\mathrm{gas}}(t)$. 
We define this time as the star formation time $t_{\mathrm{form}}$.
This condition yields $t_{\mathrm{form}} = 24.5\ \mathrm{Myr}$ in the model D1B1Z1.

We then identify dense gas clouds at this time, which are characterized as the connected region with a number density larger than $n=10^4\ \mathrm{cm^{-3}}$.
We also investigated the cases of a lower threshold density; however, we did not identify sufficiently compact clouds in these cases.

\begin{figure}[ht!]
\plotone{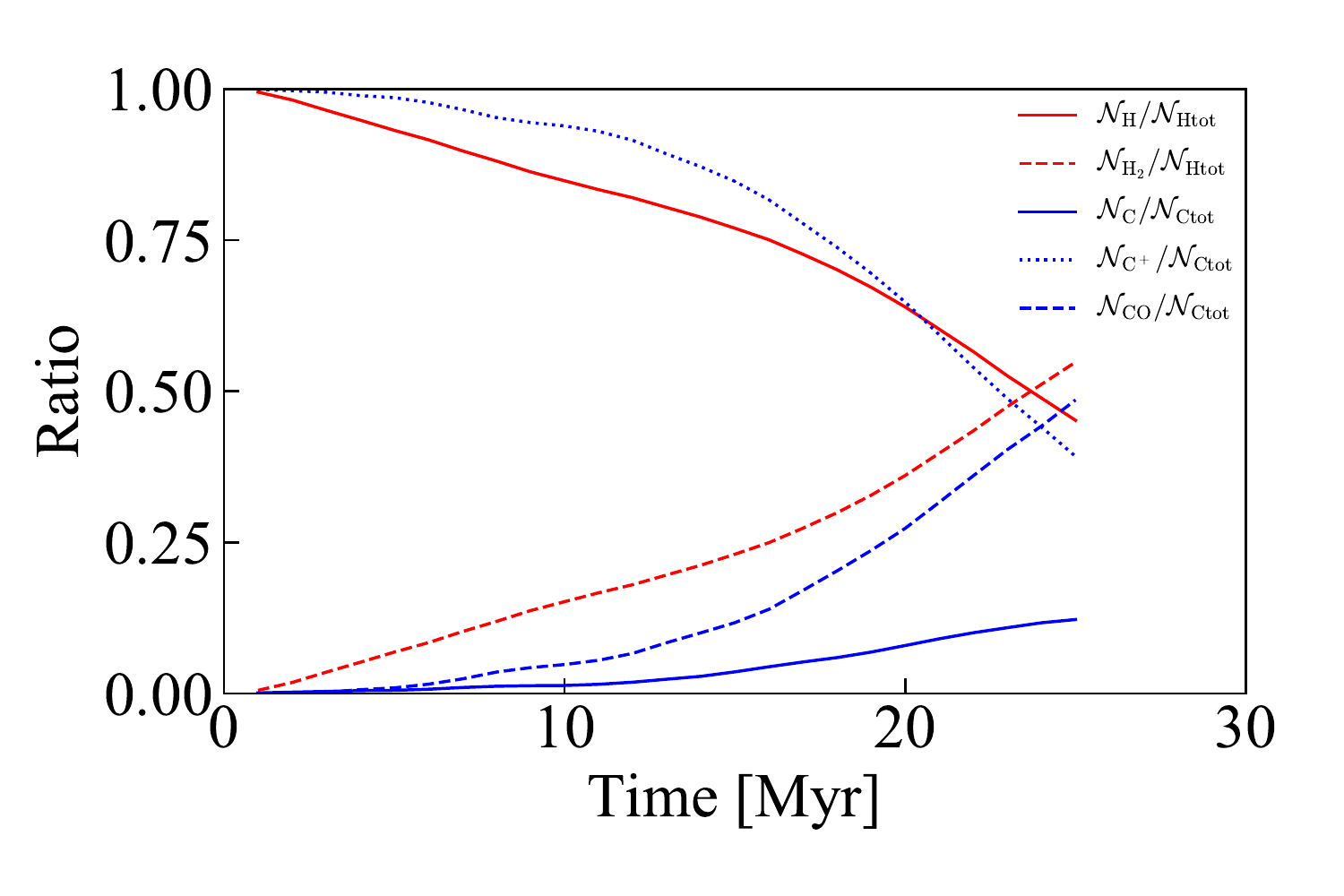}
\caption{Evolution of H fraction $\mathcal{N}_{\mathrm{H}}/\mathcal{N}_{\mathrm{Htot}}$ (red solid line), $\mathrm{H}_2$ fraction $\mathcal{N}_{\mathrm{H_2}}/\mathcal{N}_{\mathrm{Htot}}$ (red dashed line), $\mathrm{C}$ fraction $\mathcal{N}_{\mathrm{C}}/\mathcal{N}_{\mathrm{Ctot}}$ (blue solid line), $\mathrm{C^{+}}$ fraction $\mathcal{N}_{\mathrm{C^{+}}}/\mathcal{N}_{\mathrm{Ctot}}$ (blue dotted line), and CO fraction $\mathcal{N}_{\mathrm{CO}}/\mathcal{N}_{\mathrm{Ctot}}$ (blue dashed line) at the postshock region in the model D1B1Z1, where $\mathcal{N}_{\mathrm{Htot}}=\mathcal{N}_{\mathrm{H}}+ \mathcal{N}_{\mathrm{H_2}}$ and $\mathcal{N}_{\mathrm{Ctot}}=\mathcal{N}_{\mathrm{CO}}+ \mathcal{N}_{\mathrm{C}}+\mathcal{N}_{\mathrm{C^{+}}}$.  \label{fig:ratio}}
\end{figure}

Our simulations follow the formation of the $\mathrm{H_2}$ and $\mathrm{CO}$ molecules. 
The cold gas created in the shocked layer evolve into molecular gas before the cluster precursors are formed. 
 In Figure \ref{fig:ratio}, we show the evolution of H fraction $\mathcal{N}_{\mathrm{H}}/\mathcal{N}_{\mathrm{Htot}}$ (red solid line) and $\mathrm{H_2}$ fraction $\mathcal{N}_{\mathrm{H_2}}/\mathcal{N}_{\mathrm{Htot}}$ (red dashed line) at the postshock region in the model D1B1Z1, where $\mathcal{N}_{\mathrm{Htot}}=\mathcal{N}_{\mathrm{H}}+ \mathcal{N}_{\mathrm{H_2}}$.
We also represent CO fraction $\mathcal{N}_{\mathrm{CO}}/\mathcal{N}_{\mathrm{Ctot}}$ (blue dashed line),  $\mathrm{C}$ fraction $\mathcal{N}_{\mathrm{C}}/\mathcal{N}_{\mathrm{Ctot}}$ (blue solid line), and $\mathrm{C^{+}}$ fraction $\mathcal{N}_{\mathrm{C^{+}}}/\mathcal{N}_{\mathrm{Ctot}}$ (blue dotted line) at the postshock region, where $\mathcal{N}_{\mathrm{Ctot}}=\mathcal{N}_{\mathrm{CO}}+ \mathcal{N}_{\mathrm{C}}+\mathcal{N}_{\mathrm{C^{+}}}$. 
Here, $\mathcal{N}_\alpha$ means the number of $\alpha$-particles in volume $V$ ($\mathcal{N}_\alpha=\int n_\alpha\ dV$). 
In this paper, we define the molecular cloud formation time $t_{\mathrm{mol}}$ as the time when $50\%$ of the dense gas mass ($n>100\ \mathrm{cm^{-3}}$) becomes molecular. In the model D1B1Z1, we obtain $t_{\mathrm{mol}}=17.5\ \mathrm{Myr}$, indicating that the star formation time occurs $7.0\ \mathrm{Myr}$ after the molecular cloud formation time ($t_{\mathrm{form}}-t_{\mathrm{mol}}$). In Table 1, we list the values of $t_{\mathrm{form}}-t_{\mathrm{mol}}$ and $t_{\mathrm{form}}$.

\subsubsection{Case of D1B1Z1}
\begin{figure}[ht!]
\plotone{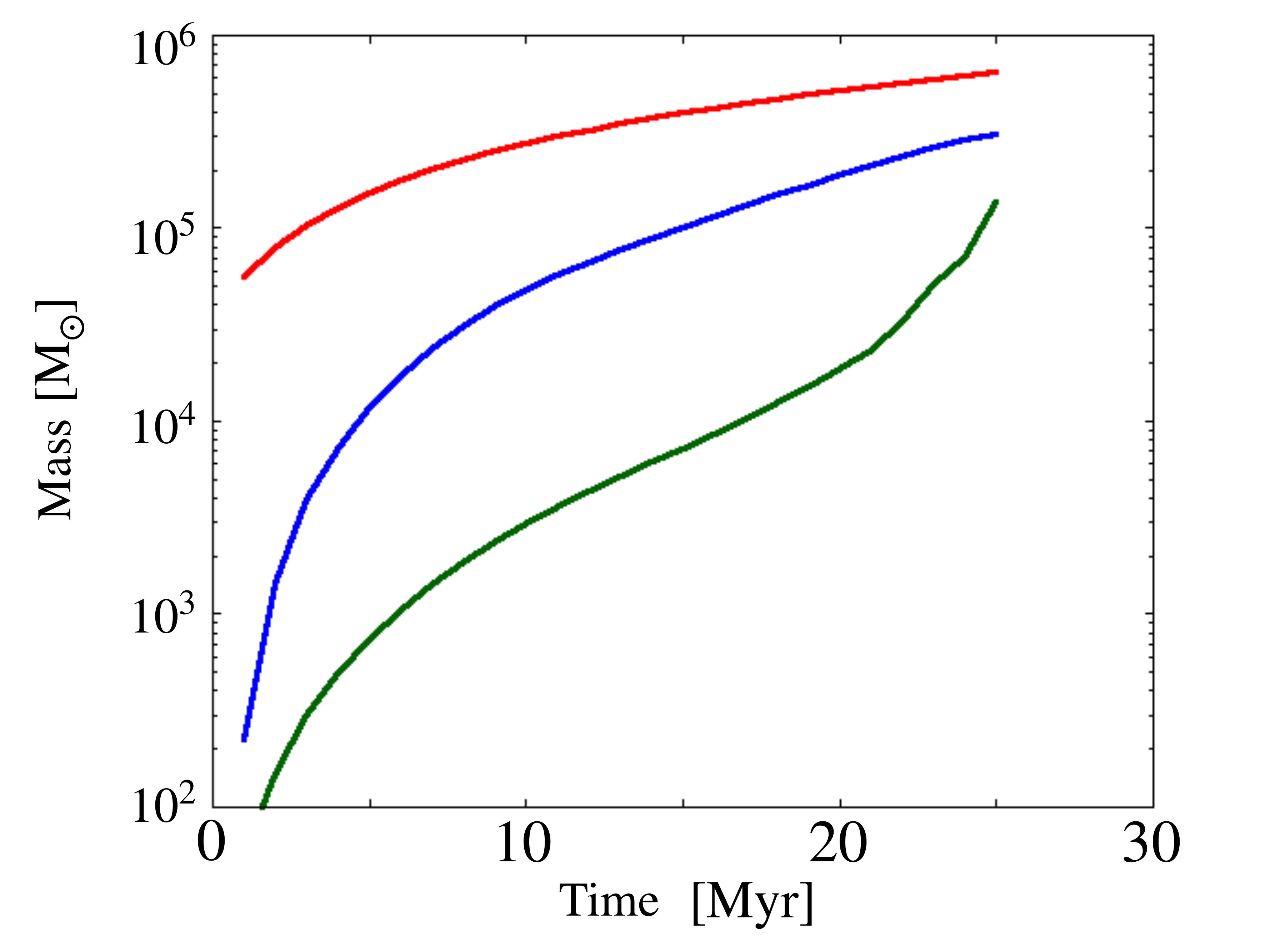}
\caption{Evolution of the total gas mass (red line), the molecular gas mass (blue line), and the mass of stars (green line) in the model D1B1Z1. \label{fig:Mass}}
\end{figure}

We apply the above identification method to the fiducial model D1B1Z1. 
Figure \ref{fig:Mass} shows the mass of the total gas (red line), the molecular gas (blue line), and the stars estimated by eq. (\ref{eq:mstar}) (green line). 
The condition $M_{\mathrm{star}}(t)/M_{{\mathrm{gas}}}(t)=0.1$ is satisfied at $t_{\mathrm{form}}=24.5\ \mathrm{Myr}$, and $t_{\mathrm{form}}-t_{\mathrm{mol}}=7.0\ \mathrm{Myr}$. 
In the following discussion, we analyze the snapshot data at this time.

\begin{figure}[ht!]
\plotone{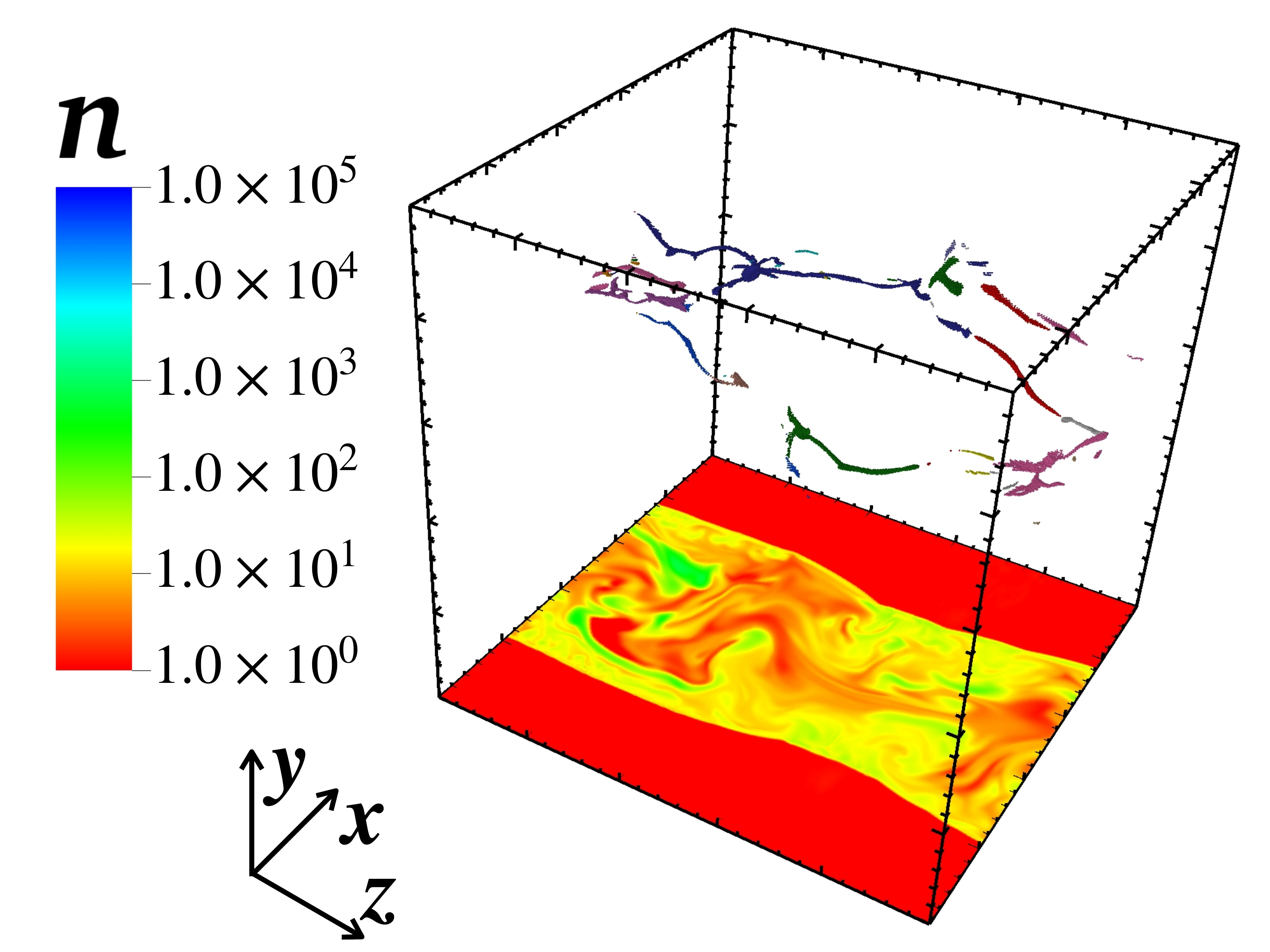}
\caption{Identified clumps in the model D1B1Z1 at the time $M_{\mathrm{star}}(t) = 0.1 M_{\mathrm{gas}}(t)$. In the box, regions of different color indicate different clumps identified by the friends-of-friends algorithm. The bottom panel represents the number density map at the $y=0$ plane. \label{fig:clumps}}
\end{figure}

\begin{figure}[ht!]
\plotone{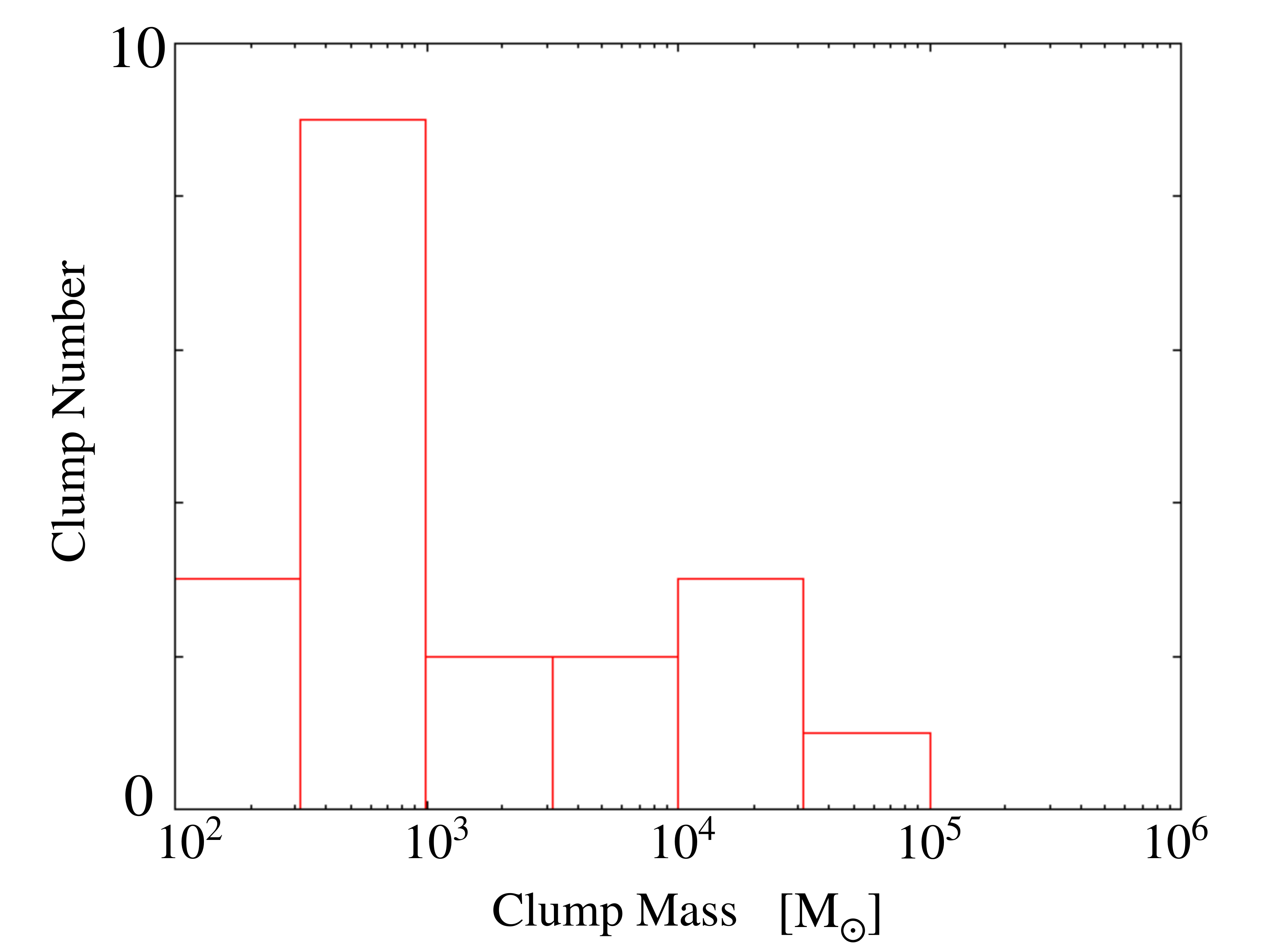}
\caption{Mass function of the identified clumps, shown in Figure \ref{fig:clumps}. The vertical axis indicates the number of clumps in the mass bin. \label{fig:masfun}}
\end{figure}

In Figure \ref{fig:clumps}, the identified dense clumps are displayed as colored regions, with different colors indicating different clouds. We identified a total of 20 dense compact clouds with $n>10^4\ \mathrm{cm^{-3}}$ and $M_{\mathrm{clump}}>10^2\ \mathrm{M_{\odot}}$ ($M_{\mathrm{clump}}$ indicates the mass of the identified clump), where the gas is already molecular. 
The corresponding mass function is plotted in Figure \ref{fig:masfun}. 
Several massive molecular clumps, with masses larger than $10^4\ \mathrm{M_{\odot}}$, are formed. The most massive clump has a mass of $M_{\mathrm{clump,max}}\sim 4\times10^4\ \mathrm{M_{\odot}}$ and a size of $L_{\mathrm{clump,max}}\sim 4\ \mathrm{pc}$, where the size is estimated by the cubic root of the volume. 
If we assume the high SFE of $30\%$, considered in the YMC forming region \citep[e.g.,][]{lada1984formation,geyer2001effect,goodwin2006gas}, these massive clumps can evolve into YMCs of mass $M_{\mathrm{YMC}}\sim 10^4\ \mathrm{M_{\odot}}$. 


\begin{figure*}[!t]
\centering
\includegraphics[clip, width=\textwidth]{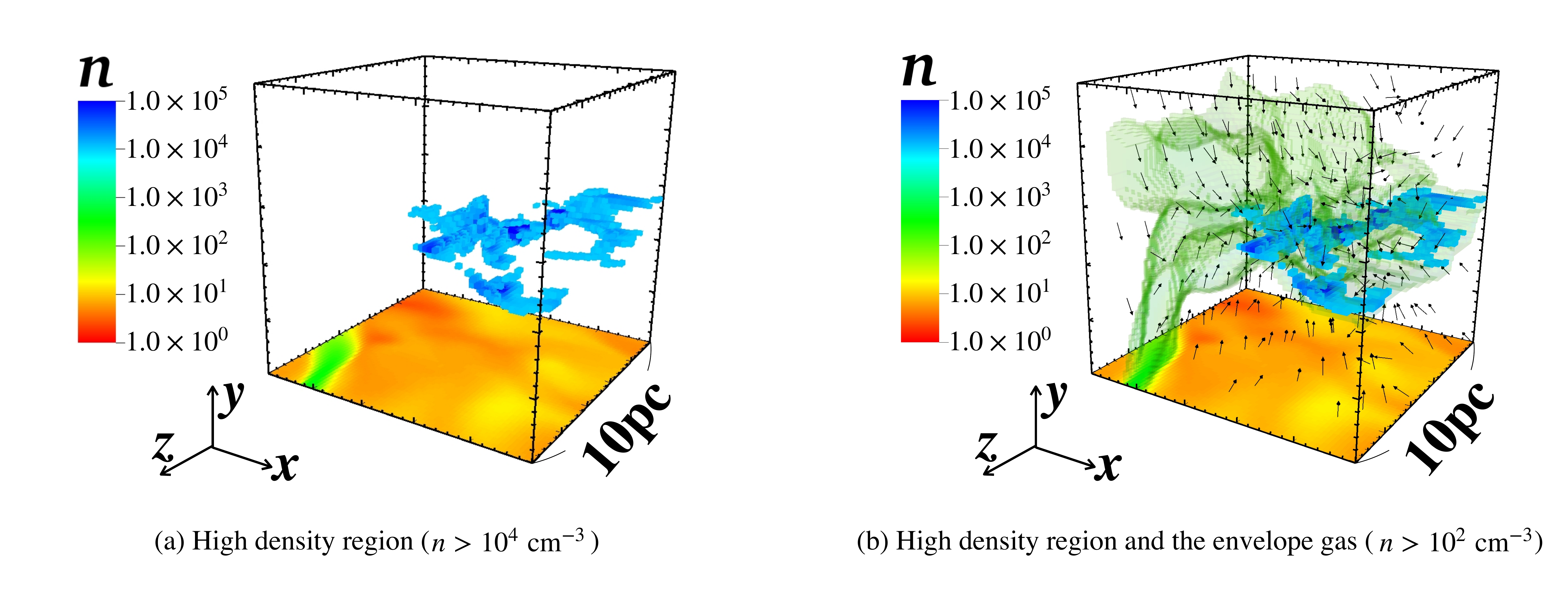}
\caption{ (a) Density structure of the most massive clump in the model D1B1Z1. The blue region represents the dense region ($n>10^4\ \mathrm{cm^{-3}}$). The center of the box matches the bottom of the gravitational potential of the most massive clump. (b) Density structure around the most massive clump in the model D1B1Z1. The blue region represents the dense region shown in (a), and the green region shows the envelope gas ($n>10^2\ \mathrm{cm^{-3}}$). The black arrows indicate the direction of the gas velocity. 
\label{fig:result-accret}}
\end{figure*}

An enlarged view of the most massive clump is shown in Figure \ref{fig:result-accret}. 
The dense region with $n>10^4\ \mathrm{cm^{-3}}$ (blue region) is surrounded by a molecular envelope with $n>10^2\ \mathrm{cm^{-3}}$ (green region). 
The arrows represent the directions of the gas velocity, indicating that the molecular envelope is accreting onto the star-forming dense clump. 
The total mass of the envelope gas (in the green region) is $\sim 2.7\times10^5\ \mathrm{M_{\odot}}$. 
Thus, the YMC precursor masses shown in Figure \ref{fig:masfun} can be enhanced with time by the accretion. 

\subsubsection{Case of D1B3Z1}
\begin{figure}[ht!]
\plotone{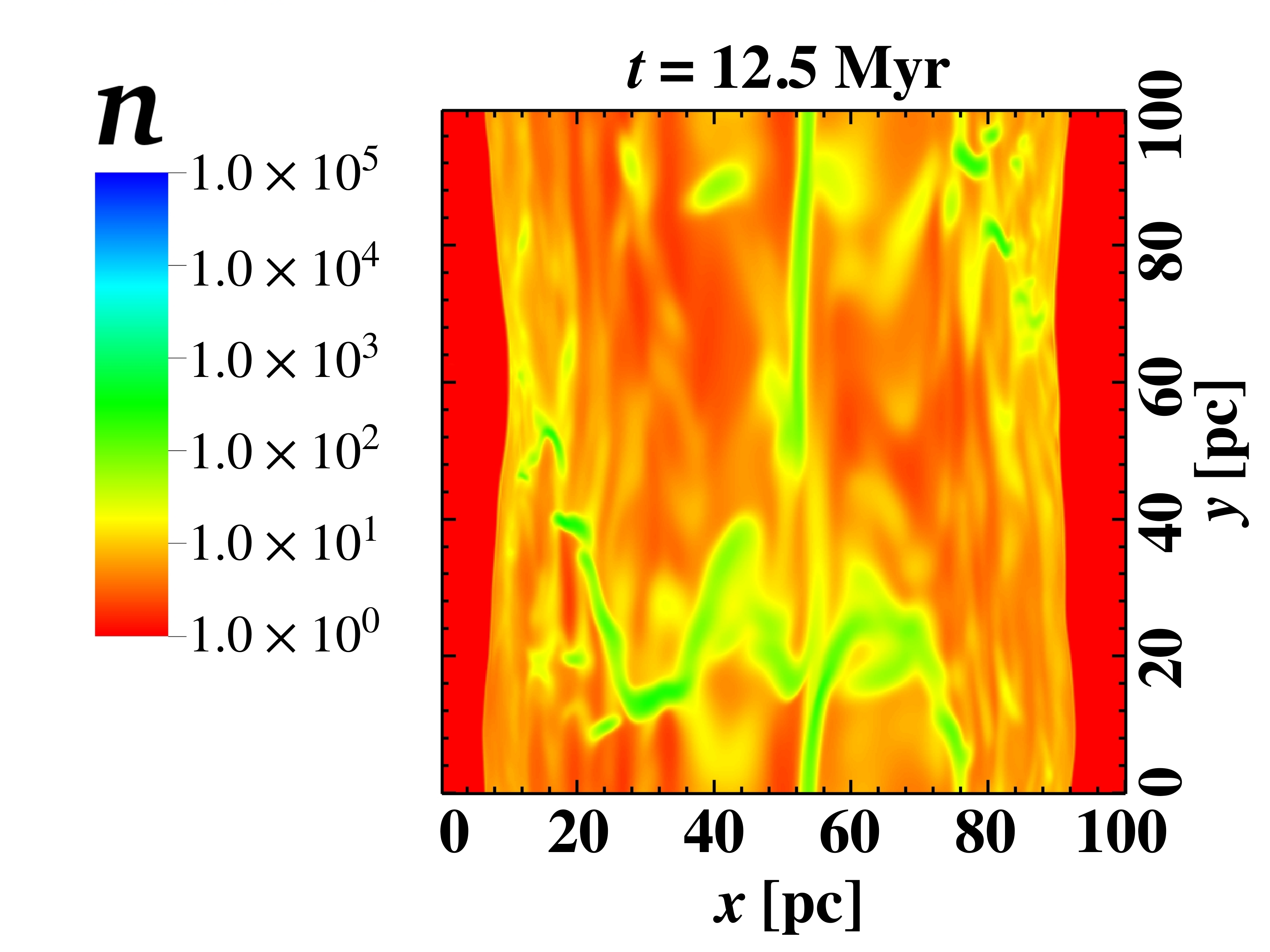}
\caption{Two-dimensional density cross section of the D1B3Z1 result at $t=12.5\  \mathrm{Myr}$.\label{fig:SB}}
\end{figure}
The density structure of the result of model D1B3Z1 is shown in Figure \ref{fig:SB}.
In this model, the initial magnetic field strength is three times larger than the fiducial model D1B1Z1. 
We attempted to identify the YMC precursor candidate clouds for this model using the method introduced in Section 3.2; however, we did not identify any high density regions with $n>10^{4}\ \mathrm{cm^{-3}}$.
From Figure \ref{fig:SB}, we can confirm that there is no dense gas with $n>10^{4}\ \mathrm{cm^{-3}}$ (blue region); however, molecular clouds with $n\sim100\ \mathrm{cm^{-3}}$ (green regions) are formed in the shocked region. 
To create a dense gas clump with $n>10^{4}\ \mathrm{cm^{-3}}$, the effect of self-gravity is necessary. 
It is widely known that the magnetized gas layer can be gravitationally unstable if the following condition is satisfied \citep{mouschovias1976note,nakano1978gravitational}: 
\begin{equation}
    \Gamma=\frac{L \bar{\rho}}{\bar{B}}\cdot 2 \pi \sqrt{G}\frac{1}{\sin \theta}>1, \label{eq:masstoflux}
\end{equation}
where $\Gamma$ is the mass to flux ratio, $\bar{\rho}$ is the postshock density of the sheet, $\bar{B}$ is the postshock magnetic field strength, $G$ is the gravitational constant, $L$ is the scale length, and $\theta$ is the angle of the magnetic field with respect to the vertical direction of the shock surface. 
In the present simulations, $\bar{\rho}$ is given by eq. (\ref{eq:shockjump-n}) and the postshock magnetic field strength can be evaluated as the balancing condition between the postshock magnetic pressure and the upstream ram pressure \citep{inoue2009two}:
\begin{equation}
\bar{B_{\perp}} \sim 35\ \mu \mathrm{G}  \left( \frac{v_{\mathrm{sh}}}{50\ \mathrm{km}\ \mathrm{s}^{-1}}\right)\left(\frac{\rho_{0}}{1\ \mathrm{m_p}\  \mathrm{cm}^{-3}}\right)^{\frac{1}{2}},\label{eq:shockjump-b}
\end{equation}
where $\mathrm{m_p}$ is the mass of a proton,  $\rho_{\mathrm{0}}$ is the density of the preshock region, $\bar{B_{\mathrm{\perp}}}$ is the magnetic field strength of the postshock region perpendicular to the converging flows, and $v_{\mathrm{sh}}$ is the shock velocity. 
Thus, we can estimate $\Gamma$ using the parameters of the D1B3Z1 model as:
\begin{equation}
    \Gamma=0.4\left(\frac{\sin \theta}{\sin 45^{\circ}}\right)^{-1}\left(\frac{L}{100 \mathrm{pc}}\right)\left(\frac{B_{0}}{3 \mu \mathrm{G}}\right)^{-1} \left(\frac{\rho_{0}}{1\ \mathrm{m_p}\ \mathrm{cm}^{-3}}\right). \label{eq:masstoflux2}
\end{equation}
Therefore, the shock compressed layer is gravitationally subcritical, which explains why the high density YMC precursor cloud is not created in this model.
If we apply $B_0=1\ \mathrm{\mu G}$, we obtain $\Gamma=1.2$, which is consistent with the result of the D1B1Z1 model, in which dense clumps are formed. 
We conclude that the YMC precursor clouds are created by the gravitational collapse of the molecular clouds formed in the shocked layer. 
Note that if we use a large box size ($L>250\ \mathrm{pc}$) or a small magnetic field angle ($\theta<17^{\circ}$), the YMC-forming clumps would be formed even when the magnetic field is stronger than $1\ \mathrm{\mu G}$.

\subsubsection{Case of D1B1Z0.2}
\begin{figure}[ht!]
\plotone{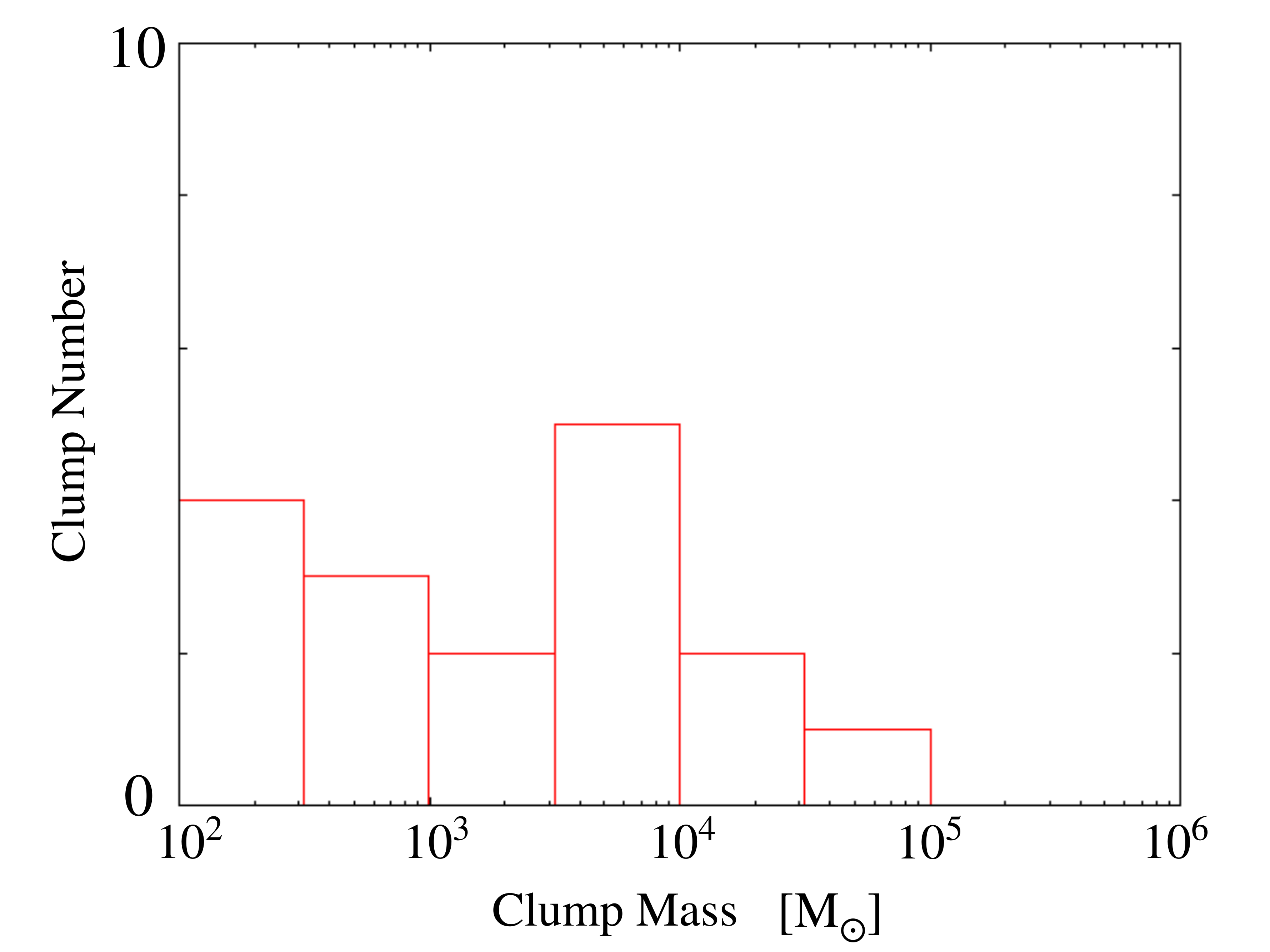}
\caption{Mass function of the clumps formed in the D1B1Z0.2 model. The vertical axis shows the number of clumps in the mass bin. \label{fig:masfun-lmtal}}
\end{figure}
We applied the same analyses for the lower metallicity model D1B1Z0.2.
The results do not show a significant difference from the fiducial model. 
The analyzed snapshot time, which satisfies $M_{\mathrm{star}}(t)/M_{\mathrm{gas}}(t)= 0.1$, is $t_{\mathrm{form}}= 24.0\ \mathrm{Myr}$ and $t_{\mathrm{form}}-t_{\mathrm{mol}}=1.0\ \mathrm{Myr}$. 
At this time, a total of 17 dense clumps are identified; the corresponding mass function is shown in Figure \ref{fig:masfun-lmtal}. 
The most massive clump has $M_{\mathrm{clump,max}}\sim 4\times 10^4\ \mathrm{M_{\odot}}$ and $L_{\mathrm{clump,max}}\sim 5\ \mathrm{pc}$, which is very similar to the result of the D1B1Z1 model. 
The reason why this model provides similar results to the fiducial model is as follows. 
The YMC precursor clumps are formed by the gravitational collapse of molecular clouds formed in the shocked layer approximately after the free-fall time estimated by eq. (\ref{eq:tff}), which is significantly longer than the cooling timescale estimated in eq. (\ref{eq:coolingtime}) for $Z=0.2$. Because the strength of the cooling affects the cloud formation timescale, but not the resulting cloud density and magnetization level of the postshock gas, the metallicity has minimal influence on the condition of eq. (\ref{eq:masstoflux}), if $t_{\mathrm{cool}}\ll t_{\mathrm{ff,sheet}}$.
This result indicates that the YMC precursor clumps can be reasonably formed, even in a low-metal environment, such as the LMC ($Z\sim 0.3\ \mathrm{Z_{\odot}}$).

\subsection{Case of D10B3Z1}
\begin{figure}[ht!]
\plotone{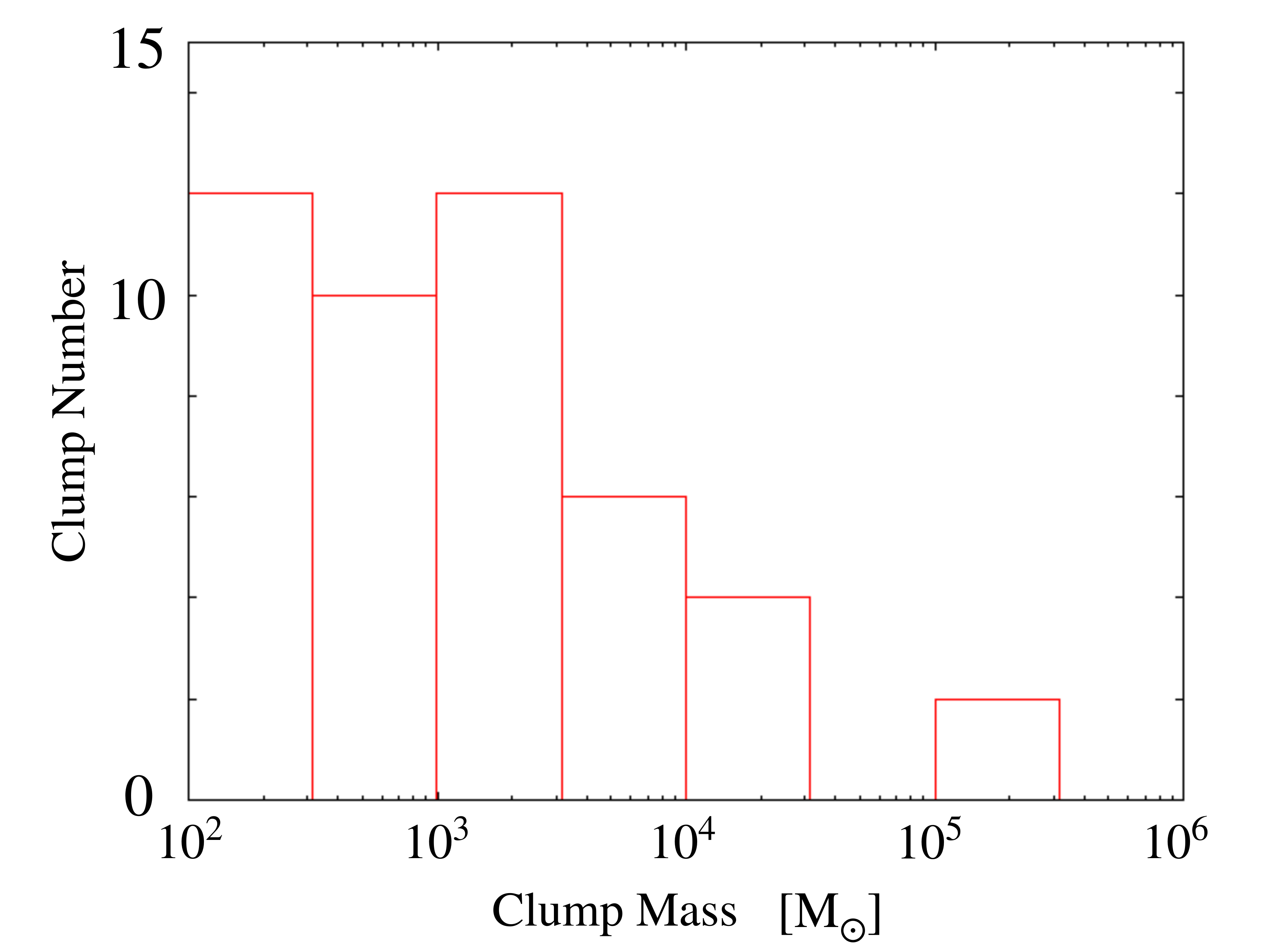}
\caption{Mass function of the clumps formed in the D10B3Z1 model. The vertical axis indicates the number of clumps in the mass bin. \label{fig:masfun-hiden}}
\end{figure}
The identified clump mass function of the high-initial-density model D10B3Z1 is shown in Figure \ref{fig:masfun-hiden}. 
The analyzed snapshot time, which satisfies $M_{\mathrm{star}}(t)/M_{\mathrm{gas}}(t)= 0.1$, is $t_{\mathrm{form}}=8.0\ \mathrm{Myr}$ and $t_{\mathrm{form}}-t_{\mathrm{mol}}=6.0\ \mathrm{Myr}$, at which $46$ dense clumps are identified. 
In this model, higher density results in a faster initiation of star formation. 
The mass of the most massive clump is $M_{\mathrm{clump,max}}\sim2\times 10^5\  \mathrm{M_{\odot}}$ and its length scale is $L_{\mathrm{clump,max}} \sim 6\ \mathrm{pc}$. 
If we apply an SFE of $30\%$, we can expect the formation of a very massive YMC of $M\sim10^5\ \mathrm{M_{\odot}}$, which is as massive as the R136 system in the LMC.

\section{DISCUSSION} \label{sec:discussion}
\subsection{The Feedback Effect of Stars}
\begin{figure}[ht!]
\plotone{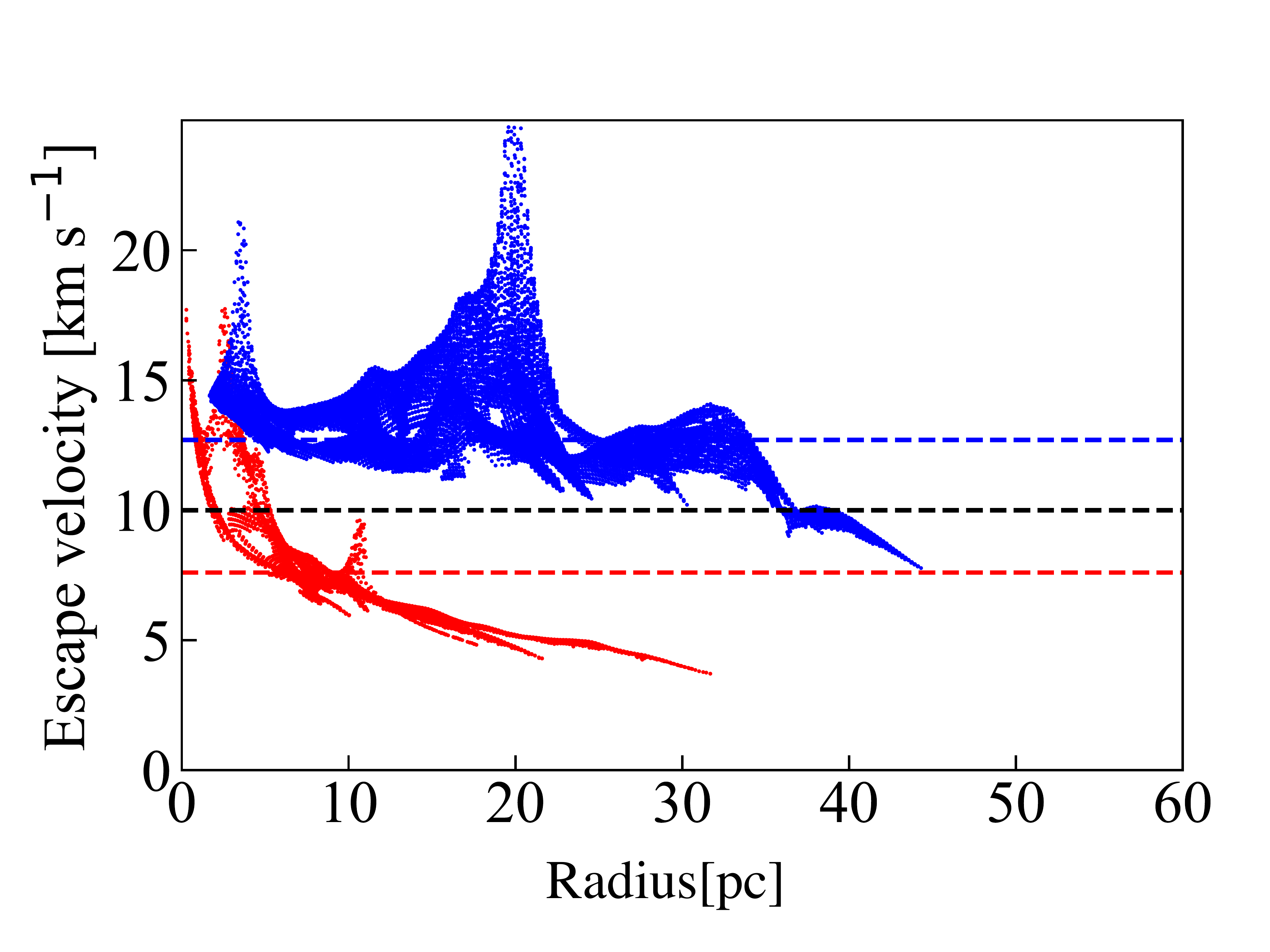}
\caption{
A scatter plot of the escape velocities of the most massive clumps identified in the model D1B1Z1 (red) and D10B3Z1 (blue).
The vertical axis (radius) means the distance from the clumps’ centers of gravity. 
The red dashed line shows the mean escape velocity ($\sim 8\ \mathrm{km\ s^{-1}}$) of the model D1B1Z1 and the blue dashed line shows the mean escape velocity ($\sim 13\ \mathrm{km\ s^{-1}}$) of the model D10B3Z1, and the black dashed line shows the typical sound speed ($\sim 10\ \mathrm{km\ s^{-1}}$) of the H{\sc ii} region. 
\label{fig:esc}}
\end{figure}
Stellar feedbacks, such as (proto)stellar outflows, radiation, and supernovae, are crucial to the regulation of star formation. 
They can blow off the gas in a YMC precursor cloud and can prevent the formation of a massive cluster, even if the parental cloud is as dense as the typical YMC. 
\cite{bressert2012find} introduced the condition under which a dense gas cloud can evolve into a YMC.
According to their study, stellar feedback is not effective enough to disintegrate and evaporate the parent cloud when the escape velocity of the YMC-forming clump is faster than the sound speed of the photo-ionized gas, which is typically $10\ \mathrm{km\ s^{-1}}$.

Here, we calculate the escape velocity employing data of the most massive clumps identified in our simulations. 
Using the density structure of the filamentary dense clump, we compute its gravitational potential $\Phi$ as an isolated system.
Then, the escape velocity is calculated as $v_{i,\mathrm{esc}} = \sqrt{2\Phi_i}$, where subscript $i$ denotes a surface cell of the dense clump.
Figure \ref{fig:esc} shows a scatter plot of the escape velocities of the most massive clumps identified in the model D1B1Z1 (red) and D10B3Z1 (blue).
The dashed lines are the mean escape velocities of the model D1B1Z1 (red) and D10B3Z1 (blue).
The average surface escape velocities are $\sim 8\ \mathrm{km\ s^{-1}}$ for the fiducial D1B1Z1 model and $\sim 13\ \mathrm{km\ s^{-1}}$ for the model D10B3Z1.
For the D10B3Z1 model, the average escape velocity is clearly greater than the typical sound speed ($\sim 10\ \mathrm{km\ s^{-1}}$) of the H{\sc ii} region. Therefore, we can naturally expect high star formation efficiency in the clump.
As for the fiducial model, the average escape velocity is smaller than the typical sound speed of the H{\sc ii} region. 
However, this does not mean the fiducial model fails to create YMCs. The mean escape velocity within $5\ \mathrm{pc}$ from the clump's center of gravity is $\sim 12\ \mathrm{km\ s^{-1}}$, where the region contains $\sim 2.5\times 10^4 \ \mathrm{M_{\odot}}$, indicating that the filamentary massive clump created in the fiducial model would also evolve into a YMC, although the edge regions are not strongly bound.
In our future work, we will test this speculation with the simulations, considering the effects of stellar feedbacks.

\subsection{Comparison with Other Colliding Flow Studies}
The gravitational collapse of molecular clouds as a consequence of colliding H{\sc i} flows has been investigated in many studies \citep[e.g.,][]{vazquez2007molecular,vazquez2009high,vazquez2011molecular,vazquez2017hierarchical,vazquez2019global,heitsch2008cooling,heitsch2008rapid,banerjee2009clump,carroll2014effects,gomez2014filaments,kortgen2015impact,zamora2018magnetic}, and it has been claimed that the global collapse triggers cluster formation \citep[e.g,][]{colin2013molecular,vazquez2017hierarchical,vazquez2019global,gonzalez2020effect}. 
In these simulations, the masses of the cluster precursor clumps were $M\sim10^3-10^4\ \mathrm{M_{\odot}}$, which are an order of magnitude smaller than the masses in our simulations. 
The reason why the previous studies did not report the formation of massive clusters of $M>10^4\ \mathrm{M_{\odot}}$ may be explained as follows: First, all the previous studies set the converging flow velocity to $\sim10\ \mathrm{km\ s^{-1}}$, which is reasonable if we consider mass accumulation by a superbubble or spiral shock. However, in our simulation, we employ flow velocities of $\sim 100\ \mathrm{km\ s^{-1}}$, which may be physically determined by the galaxy-galaxy interaction (or the escape velocity of a galaxy, such as the LMC/SMC). Second, the previous studies assumed converging flows parallel to the mean magnetic field \citep{banerjee2009clump,vazquez2011molecular,zamora2018magnetic}, which does not amplify the mean magnetic field in the shock compressed layer. In our simulations, as we have shown in Section 3.2.2, the amplification of the magnetic field, due to the finite angle, enhances the mass of the gravitationally bound objects in the shocked layer.
 Note that in the case of $\sim 10\ \mathrm{km\ s^{-1}}$ converging flows, it has been shown that magnetic field amplification prevents the formation of dense molecular clouds \citep{inoue2008two,inoue2009two}. 
In the present simulations, owing to the high flow velocity suggested by observations \citep{fukui2017formation,tsuge2019formation}, star-forming molecular clouds are created, even when the magnetic field is not parallel to the converging flows. Generally, we cannot expect the alignment of the directions of the magnetic field and highly super-Alfvénic gas flows.

\subsection{Comparison with Other Galaxy interaction Studies}
It is widely known that many YMCs were formed in interacting galaxies such as a merger galaxy \citep[e.g.,][]{ashman1992formation}. 
The simulations of star cluster formation due to a galaxy-galaxy interaction have been performed in galactic scale \citep[e.g.,][]{li2004formation,saitoh2010shocka,saitoh2010shockb}.  
They show that very massive star clusters with $\sim 10^5-10^8\ \mathrm{M_{\odot}}$ are formed in their simulations. 
In these simulations, however, they neglect the effect of the magnetic field and their mass resolution ($\sim 10^3 \ \mathrm{M_{\odot}}$) seems to be not enough to derive accurate masses of dense gas clumps and stellar clusters, although they can follow evolution of global galactic dynamics. 

\cite{elmegreen1997universal} suggested that high ambient pressure is important to form a massive compact cluster, which is consistent with recent observation \citep{tsuge2019formation2}. 
In our simulations, the large ram pressure of the colliding flows are converted into the high gas pressure (including magnetic pressure) of the shocked gas layer, where YMC precursors are formed.
Although more systematic studies are necessary, our simulations having the different ram pressures of converging flows (models D1B1Z1 and D10B3Z1) show that the larger pressure leads to a larger maximum mass of formed cluster, which is consistent with observations \citep[see e.g., Figure 17 of ][]{tsuge2019formation2}.

\section{CONCLUSION}
We investigate the YMC formation scenario triggered by the fast H{\sc i} gas collision suggested by observations \citep{fukui2017formation,tsuge2019formation}. 
We perform 3DMHD simulations, which include the effects of self-gravity, radiative cooling/heating, and chemistry. 
We obtained the following results:
\begin{itemize}
    \item[1.] The fast H{\sc i} gas collision forms massive and compact gas clumps in the postshock region. In the fiducial model D1B1Z1, the most massive clump has a mass of $\sim 4\times 10^4\ \mathrm{M_{\odot}}$ and a size of $\sim 4\ \mathrm{pc}$, which are comparable to those of the typical YMCs.
    Owing to the compact nature of the precursor clumps, the effect of stellar feedback would not be efficient, which may lead to a high SFE ($>30\%$) and resulting high-mass YMC formation.
    \item[2.] In the strong-initial-magnetic-field case (model D1B3Z1), no high-density regions with $n>10^4\ \mathrm{cm^{-3}}$ were created. This is because the postshock region remained magnetically subcritical. This result clearly shows that YMC precursor clouds are created by the gravitational collapse of molecular clouds formed in the shock compressed region. Note that if we use a larger box size ($L>250\ \mathrm{pc}$) or a smaller angle of the magnetic field with respect to the inflow ($\theta<17^{\circ}$), the YMC precursor clouds can be formed, even in the case of the D1B3Z1 model. 
    \item[3.] In a low-metal environment, such as the LMC and SMC, the YMC precursor clouds can be formed in a similar manner as the solar metallicity case, as seen in the result of the D1B1Z0.2 model. This is because the strength of the cooling affects the cloud formation timescale but not the resulting cloud density and magnetization level of the postshock gas. 
    \item[4.]In the high-initial-density model, which corresponds to H{\sc i} cloud collision, we obtained the most massive gas clump with $M>10^5\ \mathrm{M_{\odot}}$, suggesting that we can expect the formation of a very massive cluster, similar to the R136 system in the LMC. 
\end{itemize}

In this study, we did not examine the effect of star formation and stellar feedback in the simulations; however, we discussed in Section 4.1 that it would not prevent the formation of the YMC. In our future work, we will confirm our speculation using simulations considering the stellar feedback.

\acknowledgments
We thank S. Inutsuka and K. Tsuge for helpful comments and suggestions.
Numerical computations were conducted on Cray XC50 at the Center for Computational Astrophysics, National Astronomical Observatory of Japan. 
This work is supported by grant-in-aid from the Ministry of Education, Culture, Sports, Science, 
andTechnology (MEXT) of Japan, Grant No. 15K05039, No. 18H05436, and No. 20H01944 (TI).

\bibliography{sample63}{}

\begin{thebibliography}{}
\expandafter\ifx\csname natexlab\endcsname\relax\def\natexlab#1{#1}\fi
\providecommand{\url}[1]{\href{#1}{#1}}
\providecommand{\dodoi}[1]{doi:~\href{http://doi.org/#1}{\nolinkurl{#1}}}
\providecommand{\doeprint}[1]{\href{http://ascl.net/#1}{\nolinkurl{http://ascl.net/#1}}}
\providecommand{\doarXiv}[1]{\href{https://arxiv.org/abs/#1}{\nolinkurl{https://arxiv.org/abs/#1}}}

\bibitem[{Ashman \& Zepf(1992)}]{ashman1992formation}
Ashman, K.~M., \& Zepf, S.~E. 1992, \apj, 384, 50

\bibitem[{Bakes \& Tielens(1994)}]{bakes1994photoelectric}
Bakes, E., \& Tielens, A. 1994, \apj, 427, 822

\bibitem[{Ballesteros-Paredes {et~al.}(1999)Ballesteros-Paredes, Hartmann, \&
  V{\'a}zquez-Semadeni}]{ballesteros1999turbulent}
Ballesteros-Paredes, J., Hartmann, L., \& V{\'a}zquez-Semadeni, E. 1999, \apj,
  527, 285

\bibitem[{Banerjee {et~al.}(2009)Banerjee, V{\'a}zquez-Semadeni, Hennebelle, \&
  Klessen}]{banerjee2009clump}
Banerjee, R., V{\'a}zquez-Semadeni, E., Hennebelle, P., \& Klessen, R. 2009,
  \mnras, 398, 1082

\bibitem[{Bekki \& Chiba(2007)}]{Bekki2007}
Bekki, K., \& Chiba, M. 2007, \apj, 665, 1164

\bibitem[{Black \& Dalgarno(1977)}]{black1977models}
Black, J., \& Dalgarno, A. 1977, \apjs, 34, 405

\bibitem[{Bressert {et~al.}(2012)Bressert, Ginsburg, Bally, Battersby,
  Longmore, \& Testi}]{bressert2012find}
Bressert, E., Ginsburg, A., Bally, J., {et~al.} 2012, \apjl, 758, L28

\bibitem[{Carroll-Nellenback {et~al.}(2014)Carroll-Nellenback, Frank, \&
  Heitsch}]{carroll2014effects}
Carroll-Nellenback, J.~J., Frank, A., \& Heitsch, F. 2014, \apj, 790, 37

\bibitem[{Clarke(1996)}]{clarke1996consistent}
Clarke, D.~A. 1996, \apj, 457, 291

\bibitem[{Col{\'\i}n {et~al.}(2013)Col{\'\i}n, V{\'a}zquez-Semadeni, \&
  G{\'o}mez}]{colin2013molecular}
Col{\'\i}n, P., V{\'a}zquez-Semadeni, E., \& G{\'o}mez, G.~C. 2013, \mnras,
  435, 1701

\bibitem[{De~Jong {et~al.}(1980)De~Jong, Dalgarno, \&
  Boland}]{de1980hydrostatic}
De~Jong, T., Dalgarno, A., \& Boland, W. 1980, in Symposium-International
  Astronomical Union, Vol.~87, Cambridge University Press, 177--181

\bibitem[{Draine \& Bertoldi(1996)}]{draine1996structure}
Draine, B.~T., \& Bertoldi, F. 1996, arXiv preprint astro-ph/9603032

\bibitem[{Elmegreen \& Efremov(1997)}]{elmegreen1997universal}
Elmegreen, B.~G., \& Efremov, Y.~N. 1997, \apj, 480, 235

\bibitem[{Field(1965)}]{field1965thermal}
Field, G.~B. 1965, \apj, 142, 531

\bibitem[{Fujii \& Portegies~Zwart(2016)}]{fujii2016formation}
Fujii, M., \& Portegies~Zwart, S. 2016, \apj, 817, 4

\bibitem[{Fujii \& Portegies~Zwart(2015)}]{Fujii2015}
Fujii, M.~S., \& Portegies~Zwart, S.~P. 2015, Proceedings of the International
  Astronomical Union, 12, 25

\bibitem[{Fujimoto \& Noguchi(1990)}]{fujimoto1990asymmetric}
Fujimoto, M., \& Noguchi, M. 1990, \pasj, 42, 505

\bibitem[{Fujita {et~al.}(2020)Fujita, Sano, Enokiya, Hayashi, Kohno, Tsuge,
  Tachihara, Nishimura, Ohama, Yamane, {et~al.}}]{fujita2020massive}
Fujita, S., Sano, H., Enokiya, R., {et~al.} 2020, arXiv preprint
  arXiv:2003.13925

\bibitem[{Fukui {et~al.}(2017)Fukui, Tsuge, Sano, Bekki, Yozin, Tachihara, \&
  Inoue}]{fukui2017formation}
Fukui, Y., Tsuge, K., Sano, H., {et~al.} 2017, \pasj, 69, L5

\bibitem[{Fukui {et~al.}(2013)Fukui, Ohama, Hanaoka, Furukawa, Torii, Dawson,
  Mizuno, Hasegawa, Fukuda, Soga, {et~al.}}]{fukui2013molecular}
Fukui, Y., Ohama, A., Hanaoka, N., {et~al.} 2013, \apj, 780, 36

\bibitem[{Fukui {et~al.}(2016)Fukui, Torii, Ohama, Hasegawa, Hattori, Sano,
  Ohashi, Fujii, Kuwahara, Mizuno, {et~al.}}]{fukui2016two}
Fukui, Y., Torii, K., Ohama, A., {et~al.} 2016, \apj, 820, 26

\bibitem[{Furukawa {et~al.}(2009)Furukawa, Dawson, Ohama, Kawamura, Mizuno,
  Onishi, \& Fukui}]{furukawa2009molecular}
Furukawa, N., Dawson, J.~R., Ohama, A., {et~al.} 2009, \apjl, 696, L115

\bibitem[{Gaensler {et~al.}(2005)Gaensler, Haverkorn, Staveley-Smith, Dickey,
  McClure-Griffiths, Dickel, \& Wolleben}]{gaensler2005magnetic}
Gaensler, B.~M., Haverkorn, M., Staveley-Smith, L., {et~al.} 2005, Sci, 307,
  1610

\bibitem[{Geyer \& Burkert(2001)}]{geyer2001effect}
Geyer, M.~P., \& Burkert, A. 2001, \mnras, 323, 988

\bibitem[{Goldsmith \& Langer(1978)}]{goldsmith1978molecular}
Goldsmith, P., \& Langer, W. 1978, \apj, 222, 881

\bibitem[{G{\'o}mez \& V{\'a}zquez-Semadeni(2014)}]{gomez2014filaments}
G{\'o}mez, G.~C., \& V{\'a}zquez-Semadeni, E. 2014, \apj, 791, 124

\bibitem[{Gonz{\'a}lez-Samaniego \&
  Vazquez-Semadeni(2020)}]{gonzalez2020effect}
Gonz{\'a}lez-Samaniego, A., \& Vazquez-Semadeni, E. 2020, arXiv preprint
  arXiv:2003.12711

\bibitem[{Goodwin \& Bastian(2006)}]{goodwin2006gas}
Goodwin, S.~P., \& Bastian, N. 2006, \mnras, 373, 752

\bibitem[{Habing(1968)}]{habing1968interstellar}
Habing, H. 1968, BAN, 19, 421

\bibitem[{Hartmann {et~al.}(2001)Hartmann, Ballesteros-Paredes, \&
  Bergin}]{hartmann2001rapid}
Hartmann, L., Ballesteros-Paredes, J., \& Bergin, E.~A. 2001, \apj, 562, 852

\bibitem[{Heitsch {et~al.}(2005)Heitsch, Burkert, Hartmann, Slyz, \&
  Devriendt}]{heitsch2005formation}
Heitsch, F., Burkert, A., Hartmann, L.~W., Slyz, A.~D., \& Devriendt, J.~E.
  2005, \apjl, 633, L113

\bibitem[{Heitsch \& Hartmann(2008)}]{heitsch2008rapid}
Heitsch, F., \& Hartmann, L. 2008, \apj, 689, 290

\bibitem[{Heitsch {et~al.}(2008)Heitsch, Hartmann, Slyz, Devriendt, \&
  Burkert}]{heitsch2008cooling}
Heitsch, F., Hartmann, L.~W., Slyz, A.~D., Devriendt, J.~E., \& Burkert, A.
  2008, \apj, 674, 316

\bibitem[{Heitsch {et~al.}(2006)Heitsch, Slyz, Devriendt, Hartmann, \&
  Burkert}]{heitsch2006birth}
Heitsch, F., Slyz, A.~D., Devriendt, J.~E., Hartmann, L.~W., \& Burkert, A.
  2006, \apj, 648, 1052

\bibitem[{Heitsch {et~al.}(2009)Heitsch, Stone, \&
  Hartmann}]{heitsch2009effects}
Heitsch, F., Stone, J.~M., \& Hartmann, L.~W. 2009, \apj, 695, 248

\bibitem[{Hennebelle {et~al.}(2008)Hennebelle, Banerjee, V{\'a}zquez-Semadeni,
  Klessen, \& Audit}]{hennebelle2008warm}
Hennebelle, P., Banerjee, R., V{\'a}zquez-Semadeni, E., Klessen, R., \& Audit,
  E. 2008, \aap, 486, L43

\bibitem[{Hollenbach \& McKee(1979)}]{hollenbach1979molecule}
Hollenbach, D., \& McKee, C.~F. 1979, \apjs, 41, 555

\bibitem[{Hollenbach \& McKee(1989)}]{hollenbach1989molecule}
---. 1989, \apj, 342, 306

\bibitem[{Holmberg \& Flynn(2000)}]{holmberg2000local}
Holmberg, J., \& Flynn, C. 2000, \mnras, 313, 209

\bibitem[{Hosokawa \& Inutsuka(2006)}]{Hosokawa2006}
Hosokawa, T., \& Inutsuka, S. 2006, \apj, 648, L131

\bibitem[{Inoue \& Inutsuka(2008)}]{inoue2008two}
Inoue, T., \& Inutsuka, S. 2008, \apj, 687, 303

\bibitem[{Inoue \& Inutsuka(2009)}]{inoue2009two}
---. 2009, \apj, 704, 161

\bibitem[{Inoue \& Inutsuka(2012)}]{inoue2012formation}
---. 2012, \apj, 759, 35

\bibitem[{Inoue \& Inutsuka(2016)}]{Inoue2016}
---. 2016, \apj, 833, 10

\bibitem[{Inoue \& Omukai(2015)}]{Inoue2015}
Inoue, T., \& Omukai, K. 2015, \aj, 805

\bibitem[{Inutsuka {et~al.}(2015)Inutsuka, Inoue, Iwasaki, \&
  Hosokawa}]{inutsuka2015formation}
Inutsuka, S., Inoue, T., Iwasaki, K., \& Hosokawa, T. 2015, \aap, 580, A49

\bibitem[{Kim {et~al.}(1998)Kim, Staveley-Smith, Dopita, Freeman, Sault,
  Kesteven, \& McConnell}]{kim1998hi}
Kim, S., Staveley-Smith, L., Dopita, M.~A., {et~al.} 1998, \apj, 503, 674

\bibitem[{Kim {et~al.}(2003)Kim, Staveley-Smith, Dopita, Sault, Freeman, Lee,
  \& Chu}]{kim2003neutral}
---. 2003, \apjs, 148, 473

\bibitem[{K{\"o}rtgen \& Banerjee(2015)}]{kortgen2015impact}
K{\"o}rtgen, B., \& Banerjee, R. 2015, \mnras, 451, 3340

\bibitem[{Koyama \& Inutsuka(2000)}]{Koyama2000}
Koyama, H., \& Inutsuka, S. 2000, \apj, 532, 980

\bibitem[{Koyama \& Inutsuka(2001)}]{Koyama2001}
---. 2001, \apj, 564, L97

\bibitem[{Krumholz {et~al.}(2011)Krumholz, Dekel, \&
  McKee}]{krumholz2011universal}
Krumholz, M.~R., Dekel, A., \& McKee, C.~F. 2011, \apj, 745, 69

\bibitem[{Kuwahara {et~al.}(2019)Kuwahara, Torii, Mizuno, Fujita, Kohno, \&
  Fukui}]{kuwahara2019cluster}
Kuwahara, S., Torii, K., Mizuno, N., {et~al.} 2019, arXiv preprint
  arXiv:1912.00441

\bibitem[{Lada {et~al.}(1984)Lada, Margulis, \& Dearborn}]{lada1984formation}
Lada, C., Margulis, M., \& Dearborn, D. 1984, \apj, 285, 141

\bibitem[{Larson(1981)}]{larson1981turbulence}
Larson, R.~B. 1981, \mnras, 194, 809

\bibitem[{Lee {et~al.}(1996)Lee, Herbst, Pineau~des Forets, Roueff, \&
  Le~Bourlot}]{lee1996photodissociation}
Lee, H.-H., Herbst, E., Pineau~des Forets, G., Roueff, E., \& Le~Bourlot, J.
  1996, \aap, 311, 690

\bibitem[{Li {et~al.}(2004)Li, Mac~Low, \& Klessen}]{li2004formation}
Li, Y., Mac~Low, M.-M., \& Klessen, R.~S. 2004, \apjl, 614, L29

\bibitem[{Longmore {et~al.}(2014)Longmore, Kruijssen, Bastian, Bally,
  Rathborne, Testi, Stolte, Dale, Bressert, \& Alves}]{longmore2014formation}
Longmore, S.~N., Kruijssen, J.~D., Bastian, N., {et~al.} 2014, Protostars and
  Planets VI, 1, 291

\bibitem[{Millar {et~al.}(1997)Millar, Farquhar, \& Willacy}]{millar1997umist}
Millar, T., Farquhar, P., \& Willacy, K. 1997, A\&AS, 121, 139

\bibitem[{Miyama {et~al.}(1987)Miyama, Narita, \&
  Hayashi}]{miyama1987fragmentation}
Miyama, S.~M., Narita, S., \& Hayashi, C. 1987, PThPh, 78, 1273

\bibitem[{Mouschovias \& Spitzer~Jr(1976)}]{mouschovias1976note}
Mouschovias, T.~C., \& Spitzer~Jr, L. 1976, \apj, 210, 326

\bibitem[{Nakano \& Nakamura(1978)}]{nakano1978gravitational}
Nakano, T., \& Nakamura, T. 1978, \pasj, 30, 671

\bibitem[{Nelson \& Langer(1997)}]{nelson1997dynamics}
Nelson, R.~P., \& Langer, W.~D. 1997, \apj, 482, 796

\bibitem[{Ohama {et~al.}(2010)Ohama, Dawson, Furukawa, Kawamura, Moribe,
  Yamamoto, Okuda, Mizuno, Onishi, Maezawa, {et~al.}}]{ohama2010temperature}
Ohama, A., Dawson, J.~R., Furukawa, N., {et~al.} 2010, \apj, 709, 975

\bibitem[{Portegies~Zwart {et~al.}(2010)Portegies~Zwart, McMillan, \&
  Gieles}]{portegies2010young}
Portegies~Zwart, S.~F., McMillan, S.~L., \& Gieles, M. 2010, ARA\&A, 48, 431

\bibitem[{Press {et~al.}(1986)Press, Teukolsky, Vetterling, \&
  Flannery}]{press1986numerical}
Press, W.~H., Teukolsky, S.~A., Vetterling, W.~T., \& Flannery, B.~P. 1986,
  Numerical recipes in Fortran 77,  Cambridge university press New York

\bibitem[{Saitoh {et~al.}(2010{\natexlab{a}})Saitoh, Daisaka, Kokubo, Makino,
  Oakmoto, Tomisaka, Wada, \& Yoshida}]{saitoh2010shocka}
Saitoh, T., Daisaka, H., Kokubo, E., {et~al.} 2010{\natexlab{a}}, in Galaxy
  Wars: Stellar Populations and Star Formation in Interacting Galaxies, Vol.
  423, 185

\bibitem[{Saitoh {et~al.}(2010{\natexlab{b}})Saitoh, Daisaka, Kokubo, Makino,
  Okamoto, Tomisaka, Wada, \& Yoshida}]{saitoh2010shockb}
Saitoh, T.~R., Daisaka, H., Kokubo, E., {et~al.} 2010{\natexlab{b}},
  Proceedings of the International Astronomical Union, 6, 483

\bibitem[{Sano {et~al.}(1999)Sano, Inutsuka, \& Miyama}]{sano1999higher}
Sano, T., Inutsuka, S., \& Miyama, S. 1999, in Numerical Astrophysics
  (Springer), 383--386

\bibitem[{Shapiro \& Kang(1987)}]{shapiro1987hydrogen}
Shapiro, P.~R., \& Kang, H. 1987, \apj, 318, 32

\bibitem[{Spitzer(1978)}]{spitzer1978physical}
Spitzer, L. 1978, Physical Processes In The Interstellar Medium, Wiley classics
  library (Wiley \& Son)

\bibitem[{Staveley-Smith(1997)}]{staveley1997hi}
Staveley-Smith, L. 1997, PASA, 14, 111

\bibitem[{Tachihara {et~al.}(2018)Tachihara, Gratier, Sano, Tsuge, Miura,
  Muraoka, \& Fukui}]{tachihara2018triggering}
Tachihara, K., Gratier, P., Sano, H., {et~al.} 2018, \pasj, 70, S52

\bibitem[{Tielens \& Hollenbach(1985)}]{tielens1985photodissociation}
Tielens, A., \& Hollenbach, D. 1985, \apj, 291, 722

\bibitem[{Tsuge {et~al.}(2019{\natexlab{a}})Tsuge, Fukui, Tachihara, Sano,
  Tokuda, Ueda, Iono, \& Finn}]{tsuge2019formation2}
Tsuge, K., Fukui, Y., Tachihara, K., {et~al.} 2019{\natexlab{a}}, arXiv
  preprint arXiv:1909.05240

\bibitem[{Tsuge {et~al.}(2020)Tsuge, Tachihara, Fukui, Sano, Tokuda, Ueda, \&
  Iono}]{tsuge2020formation}
Tsuge, K., Tachihara, K., Fukui, Y., {et~al.} 2020, arXiv preprint
  arXiv:2005.04075

\bibitem[{Tsuge {et~al.}(2019{\natexlab{b}})Tsuge, Sano, Tachihara, Yozin,
  Bekki, Inoue, Mizuno, Kawamura, Onishi, \& Fukui}]{tsuge2019formation}
Tsuge, K., Sano, H., Tachihara, K., {et~al.} 2019{\natexlab{b}}, \apj, 871, 44

\bibitem[{Van~Leer(1997)}]{van1997flux}
Van~Leer, B. 1997, in Upwind and High-Resolution Schemes (Springer), 80--89

\bibitem[{V{\'a}zquez-Semadeni {et~al.}(2011)V{\'a}zquez-Semadeni, Banerjee,
  G{\'o}mez, Hennebelle, Duffin, \& Klessen}]{vazquez2011molecular}
V{\'a}zquez-Semadeni, E., Banerjee, R., G{\'o}mez, G.~C., {et~al.} 2011,
  \mnras, 414, 2511

\bibitem[{V{\'a}zquez-Semadeni {et~al.}(2007)V{\'a}zquez-Semadeni, G{\'o}mez,
  Jappsen, Ballesteros-Paredes, Gonz{\'a}lez, \&
  Klessen}]{vazquez2007molecular}
V{\'a}zquez-Semadeni, E., G{\'o}mez, G.~C., Jappsen, A.~K., {et~al.} 2007,
  \apj, 657, 870

\bibitem[{V{\'a}zquez-Semadeni {et~al.}(2009)V{\'a}zquez-Semadeni, G{\'o}mez,
  Jappsen, Ballesteros-Paredes, \& Klessen}]{vazquez2009high}
V{\'a}zquez-Semadeni, E., G{\'o}mez, G.~C., Jappsen, A.-K.,
  Ballesteros-Paredes, J., \& Klessen, R.~S. 2009, \apj, 707, 1023

\bibitem[{V{\'a}zquez-Semadeni {et~al.}(2017)V{\'a}zquez-Semadeni,
  Gonz{\'a}lez-Samaniego, \& Col{\'\i}n}]{vazquez2017hierarchical}
V{\'a}zquez-Semadeni, E., Gonz{\'a}lez-Samaniego, A., \& Col{\'\i}n, P. 2017,
  \mnras, 467, 1313

\bibitem[{V{\'a}zquez-Semadeni {et~al.}(2019)V{\'a}zquez-Semadeni, Palau,
  Ballesteros-Paredes, G{\'o}mez, \& Zamora-Avil{\'e}s}]{vazquez2019global}
V{\'a}zquez-Semadeni, E., Palau, A., Ballesteros-Paredes, J., G{\'o}mez, G.~C.,
  \& Zamora-Avil{\'e}s, M. 2019, \mnras, 490, 3061

\bibitem[{V{\'a}zquez-Semadeni {et~al.}(2006)V{\'a}zquez-Semadeni, Ryu, Passot,
  Gonz{\'a}lez, \& Gazol}]{vazquez2006molecular}
V{\'a}zquez-Semadeni, E., Ryu, D., Passot, T., Gonz{\'a}lez, R.~F., \& Gazol,
  A. 2006, \apj, 643, 245

\bibitem[{Wolfire {et~al.}(2003)Wolfire, McKee, Hollenbach, \&
  Tielens}]{wolfire2003neutral}
Wolfire, M.~G., McKee, C.~F., Hollenbach, D., \& Tielens, A. 2003, \apj, 587,
  278

\bibitem[{Zamora-Avil{\'e}s {et~al.}(2018)Zamora-Avil{\'e}s,
  V{\'a}zquez-Semadeni, K{\"o}rtgen, Banerjee, \&
  Hartmann}]{zamora2018magnetic}
Zamora-Avil{\'e}s, M., V{\'a}zquez-Semadeni, E., K{\"o}rtgen, B., Banerjee, R.,
  \& Hartmann, L. 2018, \mnras, 474, 4824

\end{thebibliography}
\bibliographystyle{aasjournal}

\end{document}